\documentclass[aps,prb,amsmath,amssymb,floatfix,twocolumn]{revtex4}
\usepackage{multirow}
\usepackage{bbold}
\usepackage{graphicx}
\usepackage{subfigure}
\usepackage{color}
\usepackage{hyperref}

\begin{document}
\title{Competing orders and topology in the global phase diagram of pyrochlore iridates}
\author{Pallab Goswami, Bitan Roy and Sankar Das Sarma}
\affiliation{Condensed Matter Theory Center and Joint Quantum Institute, Department of Physics, University of Maryland, College Park, Maryland 20742-4111 USA}
\date{\today}

\begin{abstract}
Strong electronic interactions and spin orbit coupling can be conducive for realizing novel broken symmetry phases supporting quasiparticles with nontrivial band topology. 227 pyrochlore iridates provide a suitable material platform for studying such emergent phenomena where both topology and competing orders play important roles. In contrast to the most members of this material class, which are thought to display ``all-in all-out" (AIAO) type magnetically ordered low-temperature insulating ground states, Pr$_2$Ir$_2$O$_7$ remains metallic while exhibiting ``spin ice" (SI) correlations at low temperatures. Additionally, this is the only 227 iridate compound, which exhibits a large anomalous Hall effect (AHE) along [1,1,1] direction below 1.5 K, without possessing any measurable magnetic moment. By focusing on the normal state of 227 iridates, described by a parabolic semimetal with quadratic band touching, we use renormalization group analysis, mean-field theory, and phenomenological Landau theory as three complementary methods to construct a global phase diagram in the presence of generic local interactions among itinerant electrons of Ir ions. While the global phase diagram supports several competing multipolar orders, motivated by the phenomenology of 227 iridates we particularly emphasize the competition between AIAO and SI orders and how it can cause a mixed phase with ``three-in one-out" (3I1O) spin configurations. 
In terms of topological properties of Weyl quasiparticles of the 3I1O state, we provide an explanation for the magnitude and the direction of the observed AHE in Pr$_2$Ir$_2$O$_7$. We propose a strain induced enhancement of the onset temperature for AHE in thin films of Pr$_2$Ir$_2$O$_7$ and additional experiments for studying competing orders in the vicinity of the metal-insulator transition. In addition to providing a theory for competing orders and magnetic properties of Pr$_2$Ir$_2$O$_7$, the theoretical framework developed in this work should also be useful for a better understanding of competing multipolar orders in other correlated materials. 
\end{abstract}

\maketitle

\section{Introduction}
The central theme of condensed matter physics is the emergence of new phases of matter in interacting many-body systems, which is succinctly captured by Anderson's famous dictum ``more is different''~\cite{Anderson1}. Magnet, superfluid and superconductor are some well known examples of emergent phases of thermodynamically large interacting systems at low temperatures, and the physical properties of such phases cannot be simply described in terms of original strongly interacting fermions or bosons. The notion of spontaneous symmetry breaking provides a unified framework for addressing these novel ordered states of matter, by allowing an efficient description in terms of a local order parameter and emergent, weakly interacting \emph{quasiparticle} excitations. Therefore, the qualitative understanding of complex phase diagrams of strongly correlated materials such as cuprates, heavy fermions and iron pnictides mainly comes from studying competing broken symmetry phases. The principle of spontaneous symmetry breaking has also been successfully applied toward explaining the emergence of massive elementary particles and color superconductivity in the context of high energy physics.

A relatively new aspect of emergence is the topological properties of quasiparticle band structure and interestingly the quadratic Hamiltonians (at mean-field level) for both gapped and gapless quasiparticles can support nontrivial topology~\cite{HasanKane,QiZhang,RyuTeo}. Consequently, even after the broken global symmetry is specified, the ground state can belong to two topologically distinct sectors, which are separated by a novel quantum phase transition (QPT). Naturally, there is tremendous current interest in finding topologically nontrivial phases in different correlated materials. The present work involves a class of iridium oxides Ln$_2$Ir$_2$O$_7$ (with Ln being a lanthanide element), also known as 227 pyrochlore iridates. By focusing on the gapless normal state of Ln$_2$Ir$_2$O$_7$, we develop a theory for the global phase diagram of these materials, showing that the system in the vicinity of metal-insulator transition manifests an intriguing confluence of competing magnetic orders and topological properties of emergent Weyl quasiparticles. 

\begin{figure}[htb]
\includegraphics[scale=0.7]{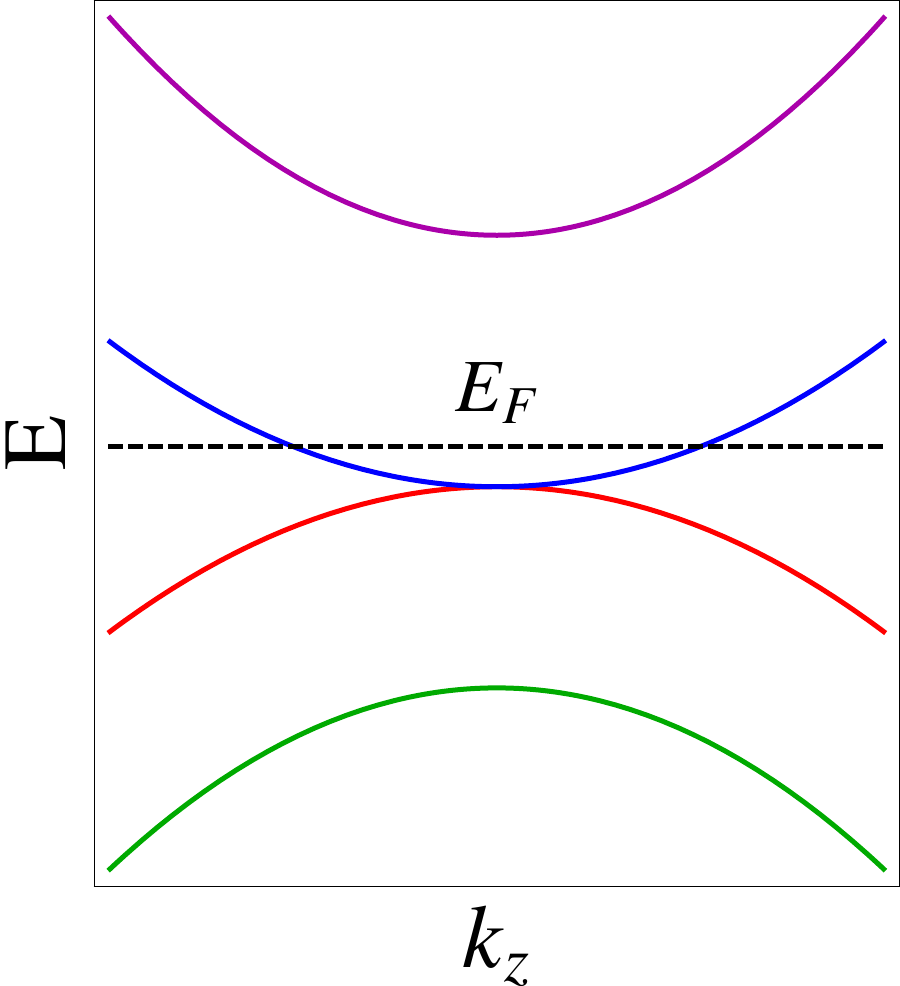}
\caption[]{Schematic representation of four Kramers degenerate bands arising from the tight-binding model for 227 pyrochlore iridates [see Eq.~(\ref{eqtb})]. At half-filling, the Fermi level $E_F$ lies within the quadratically touching bands, and the parabolic semimetal provides a suitable description of low energy physics for metallic normal state of 227 pyrochlore iridates. Our analysis will be based on this low energy subspace of quadratically touching bands, and in Fig.~\ref{nodalband} we show how they are reconstructed by symmetry breaking effects.}\label{band}
\end{figure}

\begin{figure}[htb]
\centering
\subfigure[]{
\includegraphics[scale=0.42]{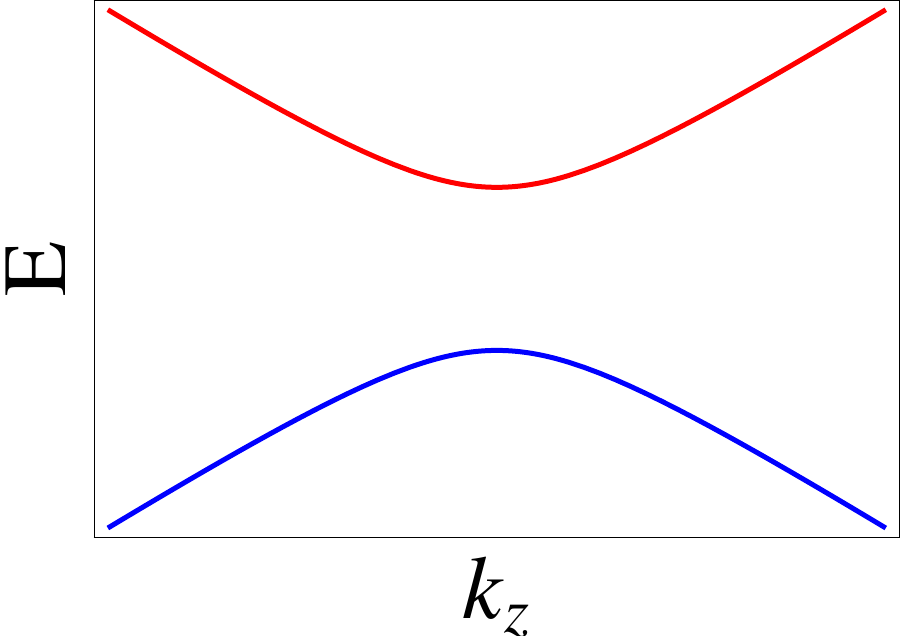}
\label{fig:bandb}
}
\subfigure[]{
\includegraphics[scale=0.42]{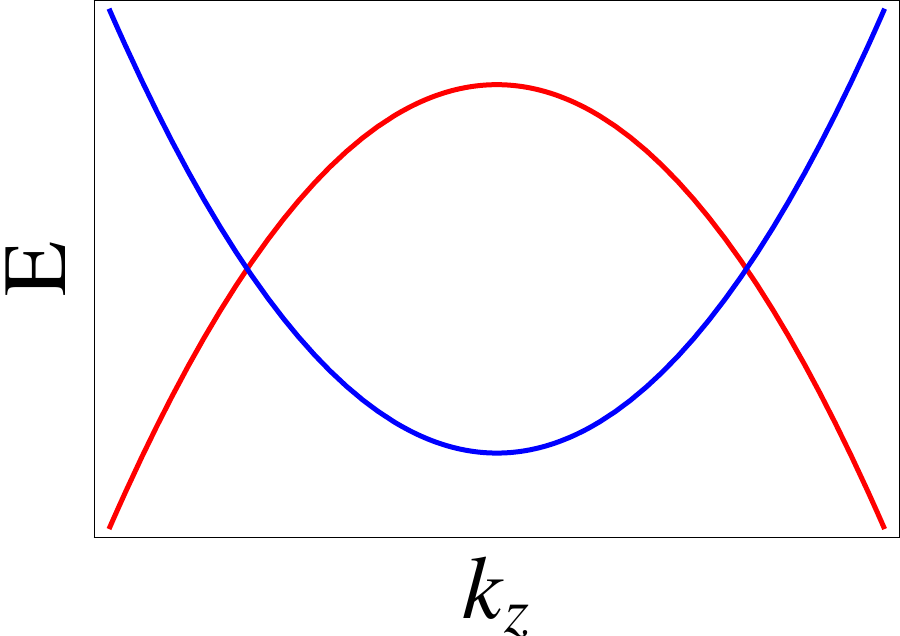}
\label{fig:bandc}
}
\subfigure[]{
\includegraphics[scale=0.42]{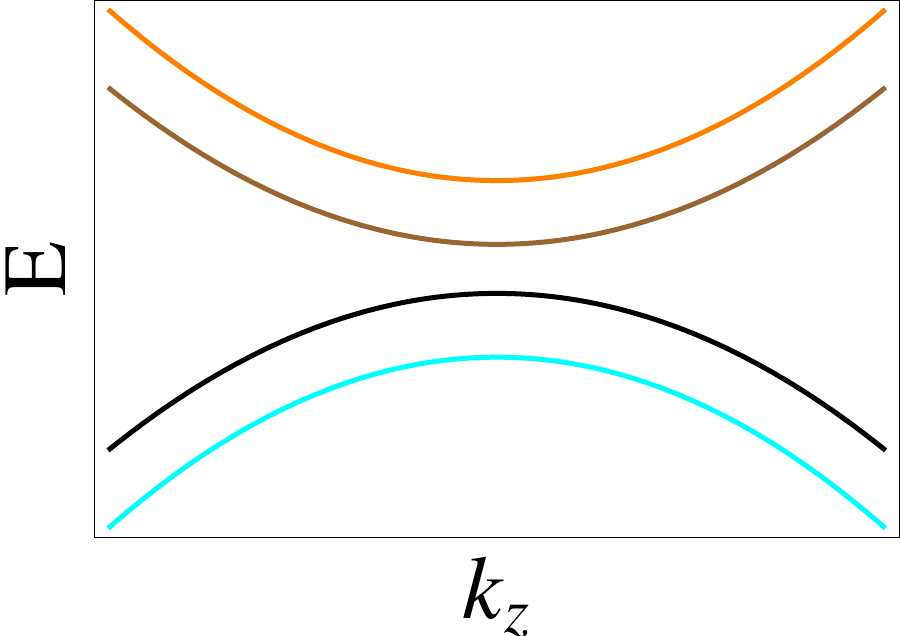}
\label{fig:bandd}
}
\subfigure[]{
\includegraphics[scale=0.42]{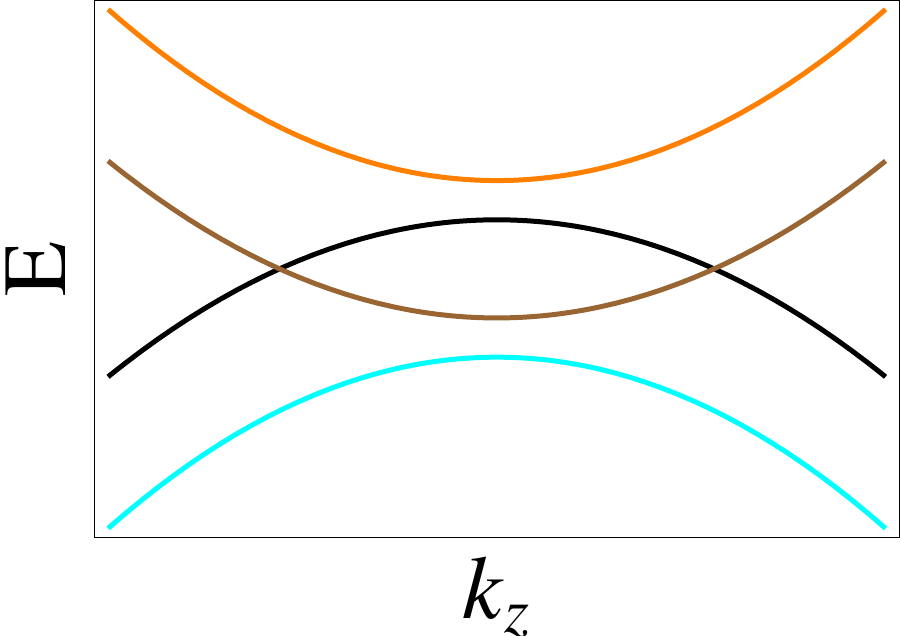}
\label{fig:bande}
}
\caption[]{Reconstruction of the quadratically touching bands shown in Fig.~\ref{band}, inside different ordered states. Time reversal symmetry preserving but rotational symmetry breaking quadrupolar or nematic orders maintain the Kramers degeneracy of conduction and valence bands and they can give rise to (a) an insulator (topological or trivial) or (b) a topological Dirac semimetal with two Dirac cones separated in the momentum space along some high symmetry direction. The Dirac semimetal phase is similar to the one observed in Na$_3$Bi and Cd$_3$As$_2$. These rotational symmetry breaking states can also be engineered by applying suitable mechanical strain. The quadratic and linear band touching points inside the parabolic and Dirac semimetal phases act as the singularities of $SU(2)$ Berry curvatures for Kramers degenerate bands and they can support large spin Hall effect. By contrast, a generic time reversal symmetry breaking order lifts the Kramers degeneracy and it can lead to (c) a trivial (or Chern) insulator or (d) a topological Weyl semimetal phase where two non-degenerate bands touch at isolated points. Since the bands in (c) and (d) are non-degenerate, the underlying Berry curvature is Abelian, which can give rise to anomalous charge and thermal Hall effects if the underlying cubic symmetry is broken. By applying an  external magnetic field or a combination of external strain and magnetic field, one can realize (c) or (d). In the context of 227 iridates ``all-in all out", ``spin ice" and ``three-in one-out" orders can give rise to Weyl semimetals. The spin configuration for these phase are showed in Fig.~\ref{Fig1}. Only ``spin ice" and ``three-in one-out" orders can support anomalous Hall effect, as they break both time reversal and cubic symmetries.}\label{nodalband}
\end{figure}

Our understanding of conventional interacting metals is guided by the renormalization group (RG) analysis of the Fermi liquid theory for low energy quasiparticles living around an underlying Fermi surface, which shows its stability against generic forward scatterings and instability toward superconductivity for infinitesimally weak attractive interactions~\cite{Shankar,Polchinski,Sachdev}. By contrast, the normal state of half-filled 227 iridates is described by a parabolic semimetal (PSM) phase~\cite{Nakatsuji6, Nakayama}, where Kramers degenerate conduction and valence bands touch quadratically at the Brillouin zone center ($\Gamma$ point) and the Fermi level lies in close proximity of the band touching point (see Fig.~\ref{band}). In the spirit of Fermi liquid theory, the RG analysis of an interacting PSM is expected to provide valuable insight into the low energy physics of 227 pyrochlore iridates. Here we present the first RG analysis of a PSM in the presence of generic local interactions, and demonstrate how sufficiently strong interactions can cause a steep competition among time reversal symmetry (TRS) preserving quadrupolar and TRS breaking dipolar and octupolar orders. With increasing strength of interactions, we find that the PSM phase can undergo continuous QPTs to different ordered states. On the other hand, in the strong interaction regime when only ordered states prevail, we construct a Landau theory to address the interplay among competing magnetic orders. Even though motivated by the phenomenology of 227 iridates (as described in the following two paragraphs), we will emphasize on TRS breaking states, the formalism outlined in this work is quite general and applicable to other correlated materials (such as half-Heusler compounds~\cite{Lin}), also displaying quadratic band touching.  Therefore, on a general ground, our work can provide valuable insight into competing multipolar orders and itinerant quantum criticality realized in many heavy fermion compounds. In Fig.~\ref{nodalband} we illustrate the reconstruction of quadratic band touching by different types of order parameter and the possible topological aspects of emergent quasiparticles. Later we will show how the topological properties of Weyl quasiparticles of a ``three-in one-out" ordered phase can  explain several enigmatic properties of Pr$_2$Ir$_2$O$_7$.

\begin{figure*}[htbp]
\centering
\subfigure[]{
\includegraphics[scale=0.36]{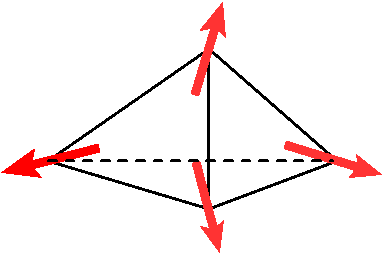}
\label{fig:1a}
}
\subfigure[]{
\includegraphics[scale=0.36]{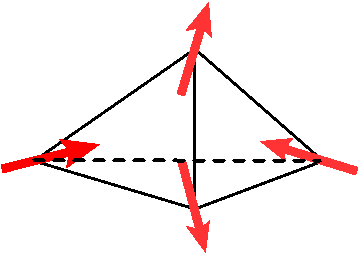}
\label{fig:1b}
}
\subfigure[]{
\includegraphics[scale=0.36]{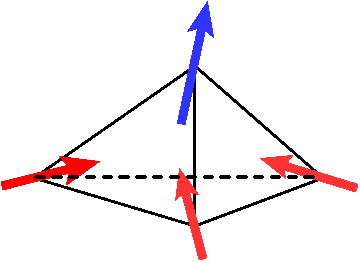}
\label{fig:1c}
}
\subfigure[]{
\includegraphics[scale=0.5]{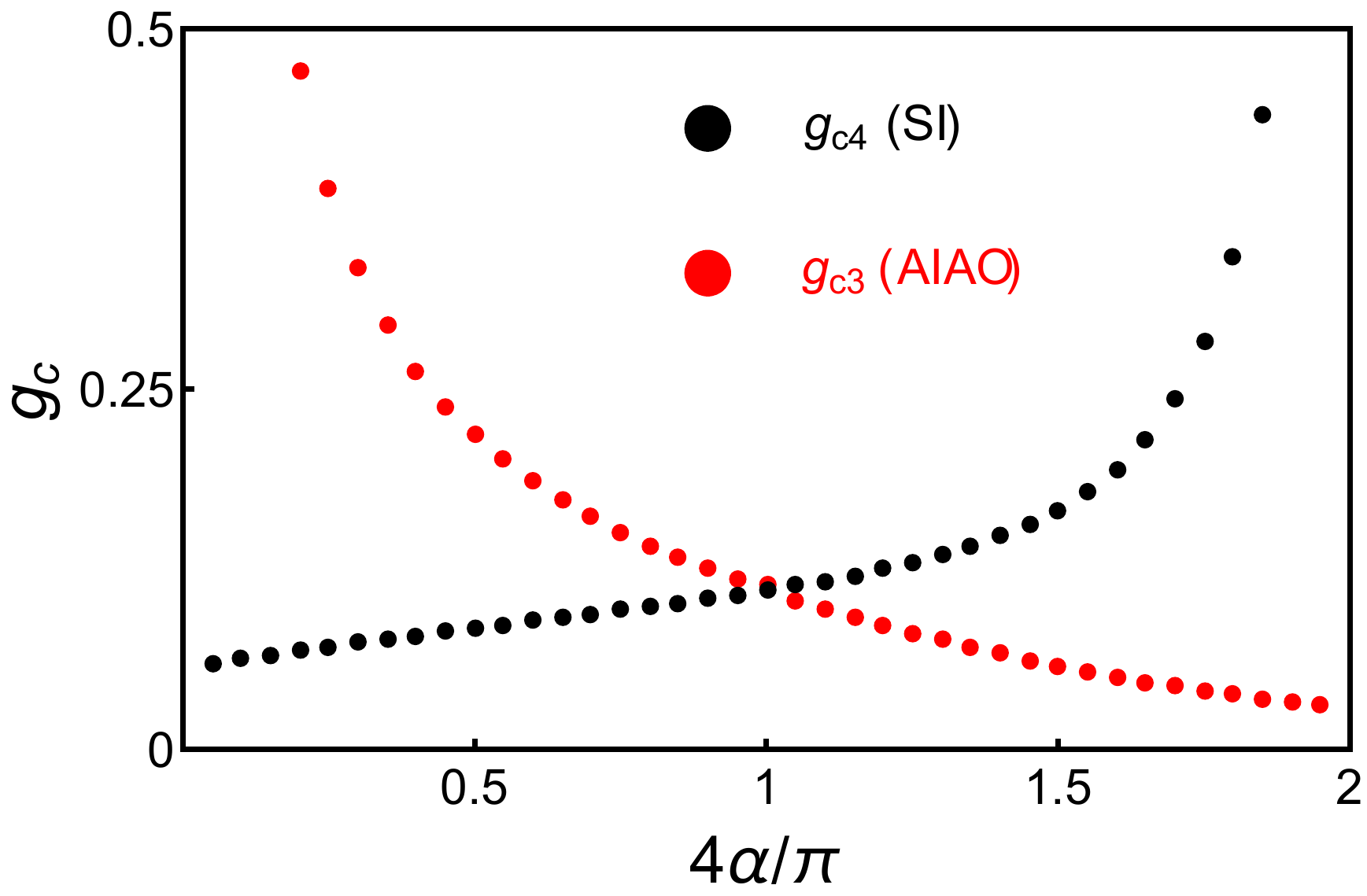}
\label{fig:1d}
}
\subfigure[]{
\includegraphics[scale=0.5]{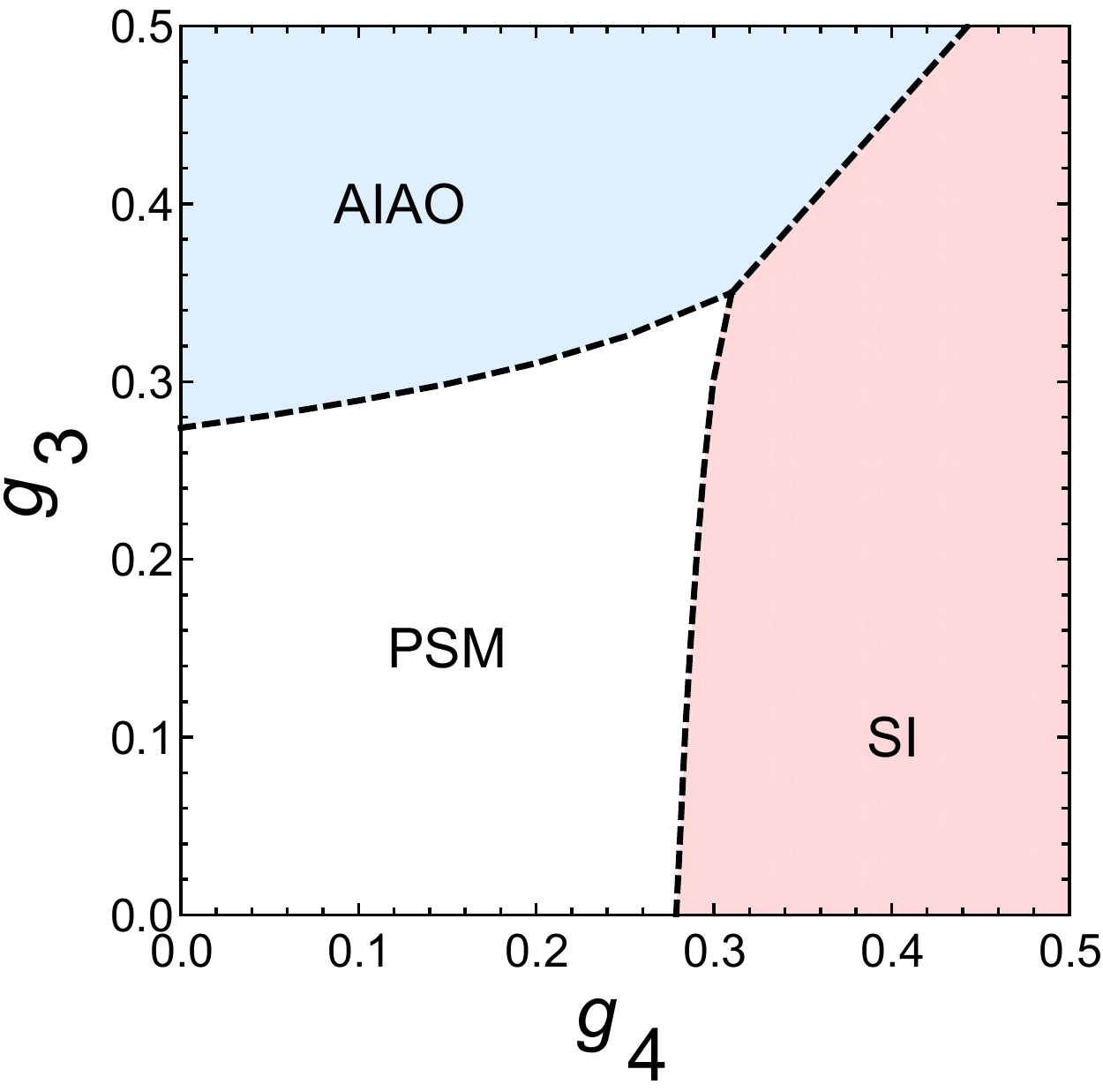}
\label{fig:1e}
}
\caption[]{(a) All out, (b) ``two-in two-out" (2I2O) and (c) ``three-in one-out" (3I1O) spin configurations on a tetrahedron. (d) Due to the vanishing quasiparticle density of states [$\rho(E) \sim |E|^{1/2}$] inside a parabolic semimetal (PSM), a broken symmetry phase appears through a continuous quantum phase transition. Corresponding mean-field critical couplings $g_{c3}$ (red) and $g_{c4}$ (black) respectively for ``all-in all-out" (AIAO) and ``spin ice" (SI) orders (in the absence of particle-hole anisotropy) are shown as functions of cubic anisotropy parameter $\alpha=\arctan(m_1/m_2)$ of the band structure [see Eq.~(\ref{Luttinger})], where $g_i$s are dimensionless interaction strengths in the AIAO and SI channels. (e) Effects of competing AIAO and SI orders on the phase diagram of a half-filled PSM in the presence of generic short range interactions, as obtained from the renormalization group calculations described in Sec.~\ref{competingorder}. How cubic symmetry finally converts the SI phase into a 3I1O ordered state, and the nature of direct transition between AIAO and 3I1O phases at strong coupling regime are addressed within a phenomenological Landau theory in Sec.~\ref{landautheory}, and the effects of direct transition between these competing states on the global phase diagram are illustrated in Fig.~\ref{Fig2}. 
}\label{Fig1}
\end{figure*}

For orientation, we first briefly discuss the current experimental results on 227 iridates. Due to comparably strong spin orbit coupling and on-site Coulomb repulsion of $5d$ Ir electrons, Ir based oxides can support metal-insulator transitions (MIT), different magnetic orders, topological phases and exotic spin liquid behaviors~\cite{Balents1,Balents2,RauKee1}. In this context, 227 pyrochlore iridates have attracted considerable experimental and theoretical interest~\cite{Takagi, Savrasov, Krempa1, Fiete1, Krempa2, Nagaosa1, Balents3, Imada2, Troyer, Fiete2, Fiete3, Ueda1, Ueda2, Drew, Ueda3, Nakatsuji1, Nakatsuji2, Nakatsuji3, Nakatsuji4, Nakatsuji5, Nakatsuji6, Nakayama, Maeno, Onoda1, Onoda2, Onoda3, Udagawa, Chen, Flint, Lee1, Rau, GChen, Nakatsuji7}. However, irrespective of the actual nature of lanthanide ion, most of the 227 iridates display insulating ground states~\cite{Takagi} and presumably an ``all in-all out" (AIAO) type of magnetic order [see Fig.~\ref{fig:1a}] at low temperatures~\cite{Savrasov, Krempa1, Krempa2, Nagaosa1, Balents3, Imada2, Troyer, Ueda1, Ueda2, Drew}. Hence, it is reasonable to assume that the correlations among $5d$ electrons of Ir ions play the dominant role in determining the global phase diagram of 227 iridates, while the size of Ln$^{3+}$ provides a suitable control parameter for tuning the strength of interactions. Consequently, by substituting for Nd$^{3+}$ with larger Pr$^{3+}$ ion one applies chemical pressure, and the MIT temperature (T$_{\mathrm{MI}}$) can be systematically suppressed to zero for Nd$_{2-2x}$Pr$_{2x}$Ir$_2$O$_7$ around a critical value $x_c \sim 0.8$~\cite{Ueda3}. A metal-insulator QPT can also be observed  by applying hydrostatic pressure on Nd$_2$Ir$_2$O$_7$, around a critical pressure $p_c \sim 5$ GPa~\cite{Ueda3}. Therefore, within the material class of 227 pyrochlore iridates, Pr$_2$Ir$_2$O$_7$ stands out as a unique system that lies in the proximity of a metal-insulator QPT, but remains metallic down to the lowest temperatures. Recent laser ARPES experiments on the normal state of both Pr$_2$Ir$_2$O$_7$ and Nd$_2$Ir$_2$O$_7$ have revealed the existence of a PSM~\cite{Nakatsuji6,Nakayama}. Next we provide a brief overview of the phenomenology of Pr$_2$Ir$_2$O$_7$, which has been a major motivating factor for the present work.

\begin{figure}[htbp]
\centering
\subfigure[]{
\includegraphics[scale=0.3]{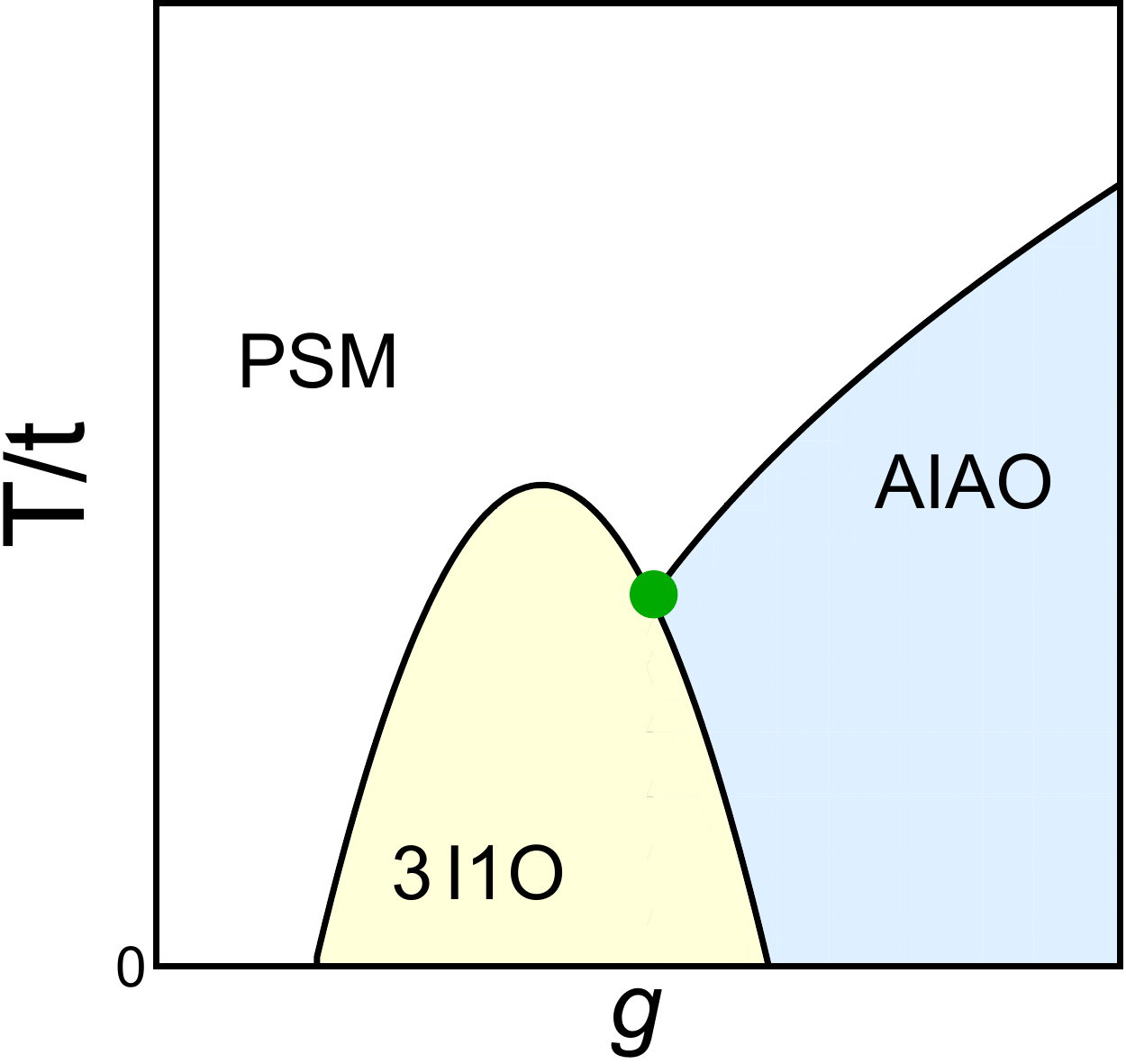}
\label{fig:2d}
}
\subfigure[]{
\includegraphics[scale=0.3]{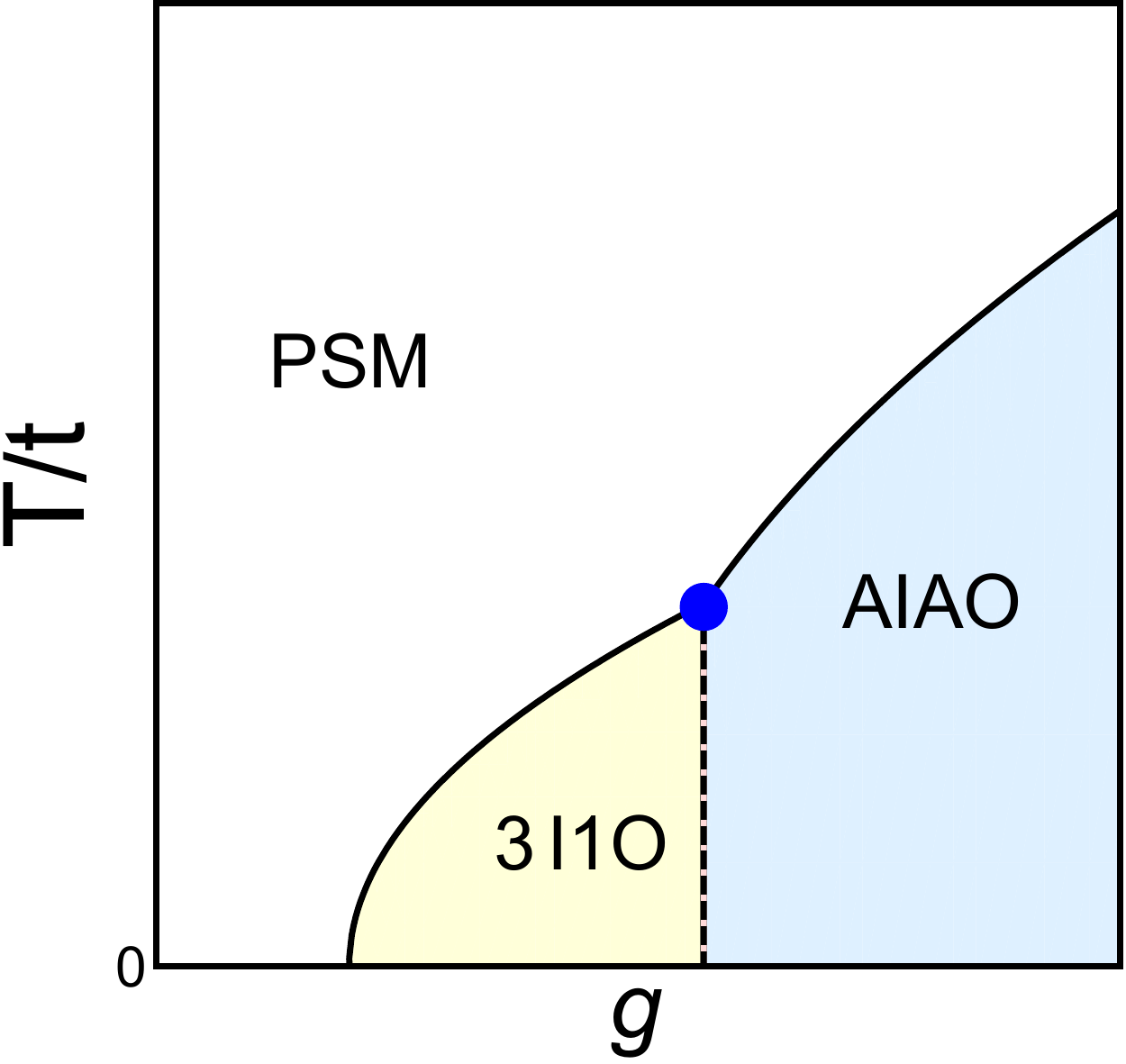}
\label{fig:2e}
}
\caption[]{Nature of the direct transition between competing ``spin ice" (SI) and ``all-in all-out" (AIAO) ordered states in the strong coupling regime of Fig.~\ref{fig:1e} and its effect on the global phase diagram of 227 iridates, as suggested by phenomenological Landau theory described in Sec.~\ref{landautheory}. Here $T$ is the temperature and $t \sim E_c$, where $E_c$ is the effective bandwidth and $g$ is a tuning parameter that depends on the strength of electronic interactions and also on external tuning parameters such as pressure, strain and magnetic field. Due to the underlying cubic symmetry, the vector order parameter for SI phase is locked along one of the eight possible [1,1,1] directions, causing an effective ``three-in one out" (3I1O) ordered state. In the presence of particle-hole anisotropy or an underlying Fermi surface in the normal state, 3I1O phase becomes an admixture of both SI and AIAO orders. (a) and (b) respectively describe two situations, when the transition between 3I1O and pure AIAO phases can be continuous and discontinuous. The green dot in (a) represents a multicritical point where three lines of continuous transitions meet. In (b) the blue dot is a bicritical point. 3I1O phase supports an anomalous Hall effect along one of the [1,1,1] directions. Since this phase is an admixture of SI and AIAO configurations, the size of Hall conductivity decreases with increasing strength of AIAO component. This prediction can be tested in Nd$_{2-2x}$Pr$_{2x}$Ir$_2$O$_7$ by decreasing $x$. Based on our theory, we expect the strength of AHE to systematically decrease as $x$ is gradually lowered from $1$, and it should vanish around the critical doping $x_c=0.8$, below which the system presumably displays only AIAO order. Based on whether the anomalous Hall conductivity vanishes continuously or discontinuously one can determine the nature of the transition between 3I1O and AIAO orders and distinguish between two possibilities shown in (a) and (b). Similar competition between these two ordered phases can also be realized by applying hydrostatic pressure on Nd$_2$Ir$_2$O$_7$.} \label{Fig2}
\end{figure}

Since the ground state of Pr is a non-Kramers doublet of $E_g$ character, the localized f electrons possess Ising dipolar moments in addition to $E_g$ quadrupolar moments~\cite{Maeno}. Experiments suggest a negative Curie-Weiss temperature $T_{CW}=-20 K$ and no apparent sign of magnetic ordering has been found~\cite{Nakatsuji1}. While a ferromagnetic exchange coupling between dipolar Ising moments of Pr (mediated by O 2p states~\cite{Onoda1,Onoda2,Onoda3}) can be reconciled with ``spin ice" (SI) (built out of ``two-in two out" configuration of Fig.~\ref{fig:1b} and its five energetically degenerate permutations) correlations observed at low temperatures~\cite{Nakatsuji2,Nakatsuji3,Nakatsuji4}, it is at odds with the observed antiferromagnetic sign of $T_{CW}$. Akin to many heavy fermion compounds, Pr$_2$Ir$_2$O$_7$ shows a resistivity minimum~\cite{Nakatsuji1}. One also finds spin freezing below 0.3 K~\cite{Nakatsuji2,Nakatsuji3}. Most strikingly, among all 227 iridates, Pr$_2$Ir$_2$O$_7$ is the only compound, which exhibits a large anomalous Hall effect (AHE) with $\sigma_{H} \sim$ 10$^{3}$ $\Omega^{-1}$ m$^{-1}$ along the $[1,1,1]$ direction in the temperature interval 0.3 K $<$ T $<$ 1.5 K without any measurable magnetic moment per Pr$^{3+}$ ion ($< 10^{-3} \mu_B$) ~\cite{Nakatsuji2,Nakatsuji3,Nakatsuji4}. In addition, it displays a metamagnetic transition only when an external magnetic field is applied along the [1,1,1] direction~\cite{Nakatsuji2,Nakatsuji3}. These experimental results have been interpreted as evidence for a metallic, chiral spin liquid phase~\cite{Nakatsuji2,Nakatsuji3,Nakatsuji4}, which have led to considerable theoretical activity~\cite{Onoda1,Onoda2,Onoda3,Udagawa,Chen,Flint,Lee1,Rau,GChen}.

Generally it has been assumed that a microscopic Kondo-Heisenberg Hamiltonian $H=H_{Ir,KE}+H_{Ir,Int}+H_{Pr}+H_{Pr-Ir}$ can describe the low energy physics of this material, where $H_{Ir,KE}$ and $H_{Ir,Int}$ respectively correspond to the kinetic energy and the inter-electron interaction for Ir. The exchange interaction between Pr local moments is denoted by $H_{Pr}$ and $H_{Ir,Pr}$ is the Kondo coupling between itinerant electrons and local moments. Most theoretical proposals have assumed the exchange interaction of Pr as the dominant energy scale and focused on exotic magnetic configurations (such as antiferroquadrupolar order of non-Kramers doublet~\cite{Onoda1,Onoda2}, Kagome ice~\cite{Udagawa}, coexisting magnetic and quadrupolar order~\cite{Lee1} and spin liquid~\cite{Rau,GChen}) for Pr$^{3+}$ local moments that can arise due to $H_{Pr}$ in the presence of strong geometrical frustration. These are ansatz ground states of a complex microscopic Hamiltonian and there is no systematic way of connecting them to the rest of the AIAO ordered iridates. On the other hand, due to the close proximity of Pr$_2$Ir$_2$O$_7$ to the MIT, we believe that $H_{Ir,Int}$ provides a comparable or larger energy scale than the exchange interaction among Pr moments. Additionally, there are general uncertainties regarding the microscopic model of a correlated material, which can be further enhanced in the proximity of MIT due to long range exchange couplings.

Thus instead of worrying about the precise microscopic model, we consider a low energy effective theory of the semimetallic normal state in the presence of generic short range interactions among itinerant electrons of Ir ions. For the metallic iridates, this should be the appropriate starting point. By employing three complementary theoretical methods: (i) RG analysis, (ii) mean-field description of ordered phases, and (iii) Landau theory of competing orders, we construct the global phase diagram of a correlated PSM. The RG analysis reveals the dominant ordering tendencies of PSM in an unbiased manner and accounts for fluctuation effects of incipient competing orders on the normal state. Once the dominant ordering susceptibility is identified, we are justified in performing mean-field calculations to describe the nature of ordered state and emergent quasiparticle excitations. Finally, based on the phenomenological Landau theory we address physical properties of the ordered states and also the transition between competing ordered phases, which provide a broader perspective of the global phase diagram (see Fig.~\ref{Fig2}). In this respect we differ from all previous theoretical works on Pr$_2$Ir$_2$O$_7$, enabling our theory to explain most of the puzzling properties of this material as arising from a competition among spin-orbit coupling, band topology (e.g. quadratic band touching), and electronic correlations. We believe that our approach can be used to study other materials where these effects dominate (e.g. half-heusler compounds, HgTe, gray-tin etc.). 

Our main results are summarized below.

\begin{enumerate}

\item Based on the representations of cubic point group, we classify all possible local order parameters for a PSM (see Table~\ref{table1}). This clearly shows that interacting PSM is an ideal itinerant system for studying competing multipolar orders, and associated itinerant quantum critical phenomena.

\item Based on the cubic symmetry, we model the generic form of local density-density interactions in terms of six independent coupling constants. By performing a RG analysis (controlled through simultaneous $\epsilon=d-2$ expansion and a large fermion flavor number $N$) of the interaction couplings and order parameter susceptibilities, we establish the competition between metallic SI (supporting ``2-in, 2-out" (2I2O) spin structures on Ir tetrahedron) and AIAO orders. The expectation values of the spin operator for Ir electrons due to AIAO and SI (one of the six possible 2I2O configurations) orders are respectively shown in Fig.~\ref{fig:1a} and Fig.~\ref{fig:1b}. The critical coupling strengths for SI and AIAO orders at mean-field level are shown in Fig.~\ref{fig:1d}. The effects of these two competing orders on the phase diagram of an interacting PSM as suggested by the RG analysis are shown in Fig.~\ref{fig:1e}. Notice that in the presence of suitable local interactions, the PSM can undergo continuous quantum phase transitions into either of these two ordered states.  

\item Within our model of generic short range interactions, there are three distinct coupling constants in the magnetic channel [$g_3$, $g_4$ and $g_5$ in Eq.~(\ref{genericint})], capturing the effects of magnetic anisotropy allowed by the cubic symmetry. Inside the dominant ordered states, such as AIAO and SI phases, the magnetic moments of Ir ions lock along the diagonals of a tetrahedron. This does not happen as a consequence of any assumed local [1,1,1] or Ising anisotropy for Ir ions. Within the extended model of interactions, there are strong coupling, renormalization group fixed points, which describe magnetically ordered states with locked Ir moments along the diagonals of a tetrahedron. Therefore, the locking of Ir moments is entirely a consequence of spontaneous symmetry breaking due to sufficiently strong electronic interactions.  

\item When interactions are so strong that a PSM is absent as a zero temperature phase, we study the nature of transition between SI and AIAO ordered states by using a phenomenological Landau theory. For a magnetic phase with vector order parameter, cubic crystal symmetry allows for two distinct quartic couplings in the magnetic free energy [as described by $u^{\prime}_{1}$ and $u^{\prime \prime}_{1}$ in Eq.~(\ref{lfph})]. At low temperatures, such terms cause locking of the vector order parameter along one of the eight possible [1,1,1] directions. Since our SI phase is described by a vector order parameter, we show that the underlying cubic environment locks the SI order parameter or magnetic moment along one of the eight possible [1,1,1] directions, thus converting the SI phase into a 3I1O ordered state at low temperatures. We will emphasize that within our theoretical analysis, the emergent 3I1O order has nothing to do with local Ising anisotropy of Pr ions. ``3-in, 1-out" (3I1O) configuration shown in Fig.~\ref{fig:1c} will play an important role in the low energy physics of Pr$_2$Ir$_2$O$_7$ and will be discussed below. The nature of transition between AIAO and 3I1O phases is illustrated in Fig.~\ref{Fig2}. For a generic model of particle-hole anisotropic PSM, we find that 3I1O order supports an admixture of both SI and AIAO spin configurations on the Ir tetrahedron. 

\item The 3I1O order gives rise to a Weyl metal with two Weyl nodes (as shown in Fig.~\ref{fig:bande}) separated along one of the eight possible [1,1,1] directions. Such a phase can support a large AHE without any appreciable magnetic moment for Ir$^{4+}$ and Pr$^{3+}$ ions, as observed in Pr$_2$Ir$_2$O$_7$ at low temperatures. Due to the quadratic dispersion relation in the parent PSM phase, the order parameter amplitude ($|\mathbf{M}|$), the mean-field transition temperature or the onset temperature of AHE ($T_H$) and the anomalous Hall conductivity ($\sigma_H$) inside the 3I1O ordered phase are related to each other according to $|\mathbf{M}| \sim k_B T_H \sim \sigma^2_H$. 

\item Inside the metallic 3I1O phase, we find that the constituting SI and AIAO components couple to the fluctuating $T_{2g}$ quadrupolar order parameter, causing a small nematicity for the system. Consequently, we show that an externally applied strain along [1,1,1] direction can induce a uniform coupling between SI and AIAO orders, and how it can enhance $T_H$ and $\sigma_H$, which should be observable in experiments on thin films of Pr$_2$Ir$_2$O$_7$. We also discuss additional experimental signatures of the competing order parameters.

\end{enumerate}

Rest of the paper is organized as follows. In Sec.~\ref{tight-binding} we discuss the tight-binding model and the effective low energy band structure of 5d itinerant electrons of Ir. The Sec.~\ref{competingorder} is devoted to the discussion of competing order parameters and their coupling to gapless fermions. Here we provide the RG analysis of a parabolic semimetal in the presence of generic local interactions. The fermion spectra, nodal topology, and the possibility of AHE in various magnetic ground states are discussed Sec.~\ref{AHE}. The explanation of AHE in Pr$_2$Ir$_2$O$_7$ in terms of Weyl metal phase induced by the 3I1O order is provided in Sec.~\ref{section-pr}. Based on a phenomenological Landau theory we address the competition among dominant magnetic orders for 227 pyrochlore iridates in Sec.~\ref{landautheory}. In Sec.~\ref{strain-magfield} we consider the effects of external strain and magnetic fields on the competing orders, and discuss various experimental ramifications of our theory. We summarize our findings in Sec.~\ref{conclusion}. The technical aspects of deriving the Luttinger model beginning with a tight-binding Hamiltonian and the coupling between itinerant fermions and magnetic order parameters are respectively provided in Appendix~\ref{sec:appendix-A} and Appendix~\ref{sec:appendix-B}. A detailed description of the RG analysis can be found in Appendix~\ref{sec:appendix-RG}.

\section{Luttinger model}\label{tight-binding}

The band structure of 5d electrons of Ir$^{4+}$ ions can be described by the following tight-binding Hamiltonian~\cite{Imada1,Krempa2}
\begin{eqnarray}
H_{tb}&=&\sum_{\langle ij \rangle} \; \sum_{s,s^\prime} \; c^\dagger_{i,s} \left[ 2t_1 \; \delta_{s,s^\prime} + i \xi \; \boldsymbol \sigma_{s,s^\prime} \cdot \mathbf{e}_{ij} \right]c_{j,s^\prime} \nonumber \\
&+&2t_2 \; \sum_{\langle \langle ij \rangle \rangle} \; \sum_{s,s^\prime} \; \delta_{s,s^\prime} \; c^\dagger_{i,s} c_{j,s^\prime},\label{eqtb}
\end{eqnarray}
where $t_1$, $t_2$ respectively denote spin independent hopping strengths among nearest neighbors and next nearest neighbors, whereas $\xi$ is the nearest neighbor spin dependent hopping amplitude. The fermion annihilation operator at site $i$ with spin projection $s=\pm 1/2$ is $c_{i,s}$, and $\boldsymbol \sigma$ are three Pauli matrices. The vectors $\mathbf{e}_{ij}=(\mathbf{b}_{ij} \times \mathbf{d}_{ij})/|\mathbf{b}_{ij} \times \mathbf{d}_{ij}|$, where $\mathbf{b}_{ij}$ is the vectors directed from the center of a tetrahedron to the midpoint of the bond $\langle ij \rangle$, and $\mathbf{d}_{ij}$ is the vector directed from site $\mathbf{r}_i$ to site $\mathbf{r}_j$.

The above tight-binding model gives rise to two high energy conduction and valence bands with their minimum and maximum respectively located at reference energies $E_+=6(t_1+2t_2)$, $E_-=-2(t_1+2t_2)-4\sqrt{2} \xi$. In addition, a set of Kramers degenerate conduction and valence bands touch quadratically at reference energy $E_0=-2\left(t_1+2t_2 \right)+2 \sqrt{2} \xi$, as schematically shown in Fig.~\ref{band}. At half-filling the Fermi level lies inside the quadratically touching bands if all the tight-binding parameters are chosen to be positive or negative. However the band structure of 227 iridates are qualitatively reproduced only for the positive parametrization. Within the low energy subspace of quadratically touching bands the effective Hamiltonian is described by the Luttinger model~\cite{luttinger}
\begin{equation}
\hat{H}_{L}= E_0(\mathbf{k}) \mathbb{1}-\frac{\hbar^2}{2m_1} \; \sum_{j=1}^{3} \; d_j(\mathbf{k}) \Gamma_j-\frac{\hbar^2}{2m_2} \; \sum_{j=4}^{5} \; d_j(\mathbf{k}) \Gamma_j, \label{Luttinger}
\end{equation}
where $E_0(\mathbf{k})=E_0+ \hbar^2 \mathbf{k}^2/(2m_0)$ and
\begin{eqnarray}
d_a(\mathbf{k})&=&-\frac{\sqrt{3}}{2} |\epsilon_{abc}| k_b k_c, \: d_4(\mathbf{k})=-\frac{\sqrt{3}}{2} (k^2_1-k^2_2), \nonumber \\
 \; d_5(\mathbf{k})&=&-\frac{1}{2} (2k^2_3-k^2_1-k^2_2).
\end{eqnarray}
The latin indices $a,b,c$ take values out of $1,2,3$. The five mutually anticommuting $\Gamma$ matrices can be defined in terms of spin-$3/2$ matrices $J_1$, $J_2$ and $J_3$ according to~\cite{murakami}
\begin{eqnarray}
\Gamma_a&=&\frac{1}{\sqrt{3}} |\epsilon_{abc}| J_b J_c, \: \Gamma_4=\frac{1}{\sqrt{3}} (J^2_1-J^2_2), \nonumber \\
 \; \Gamma_5&=&\frac{1}{3} (2J^2_3-J^2_1-J^2_2).
\end{eqnarray}
A detailed derivation of the Luttinger model is presented in Appendix~\ref{sec:appendix-A}. Generally the effective mass parameters $m_0 \neq m_1 \neq m_2$, and the fermion spectra are given by $E_{n,s}(\mathbf{k})=E_0(\mathbf{k}) + \mathrm{sgn}(n) \; \epsilon(\mathbf{k})$ with
\begin{equation}
\epsilon(\mathbf{k})=\frac{\hbar^2}{2 m_p} \; \sqrt{\sin^2 \alpha \sum_{j=4}^{5} d^2_j(\mathbf{k})+\cos^2 \alpha \sum_{j=1}^{3} d^2_j(\mathbf{k})},
\end{equation}
where $n=\pm$ respectively correspond to conduction and valence bands, and the independence of the dispersion on $s=\pm$ represents the Kramers degeneracy. We have also defined $\tan \alpha = m_1/m_2$, with $m_p=m_1 m_2/\sqrt{m^2_1+m^2_2}$. In terms of the tight-binding parameters of Eq.~(\ref{eqtb}), the effective masses can be expressed as
\begin{eqnarray}
m_0=\frac{3\hbar^2 a^{-2}}{4(t_1+6t_2)}, m_1=\frac{3\hbar^2 a^{-2}}{4(t_1-2t_2)},
m_2=\frac{3\hbar^2 a^{-2}}{4(t_1-6t_2)}, \nonumber
\end{eqnarray}
where $a$ is the lattice spacing. When $t_2/t_1>0$, we find $m_1<m_2$, $\alpha<\pi/4$ and vice versa. The five component vector $\mathbf{d}(\mathbf{k})$ describes the explicit form of spin orbital locking at a given wavevector $\mathbf{k}$. Since all components of $\mathbf{d}(\mathbf{k})$ vanish at the $\Gamma$ point, four degenerate states possess an SU(4) symmetry, which at any finite $\mathbf{k}$ is reduced to $SU(2) \otimes SU(2) \sim SO(4)$ symmetry of the Kramers degenerate conduction and valence bands~\cite{murakami}. We note that the Luttinger model is invariant under the TRS operation defined by $\mathbf{k} \to -\mathbf{k}$ and $\Psi_{\mathbf{k}} \to i\Gamma_3\Gamma_1 \Psi_{-\mathbf{k}}$.

A PSM phase is also realized in several weakly correlated gapless semiconductors HgTe, grey-Sn [see Ref.~\onlinecite{Nimtz}], and strongly correlated half-heusler compounds~\cite{Lin}. It is important to notice that the PSM describes a \emph{fermionic quantum critical} system with dynamic scaling exponent $z=2$, as $\epsilon(\mathbf{k}) \sim |\mathbf{k}|^2$. Consequently, all the critical (singular power law) thermodynamic and transport properties of a PSM are entirely determined by $z=2$ and the spatial dimensionality $d=3$. Since the density of states for a three dimensional PSM vanishes as $\rho(E) \sim E^{d/z-1}=E^{1/2}$, we can immediately infer the following scaling behaviors: specific heat $C_v \sim T^{3/2}$, compressibility $\kappa \sim T^{1/2}$, the dynamic interband conductivity $\sigma(\omega) \sim \omega^{1/2}$. In correlated materials supporting broken symmetry phases at low temperatures, such scaling behaviors can only be seen above the ordering temperatures. Since Pr$_2$Ir$_2$O$_7$ and Nd$_2$Ir$_2$O$_7$ lie very close to the quantum MIT, the low temperature signatures of PSM have been clearly observed only in these two materials.

\section{Competing orders and RG analysis}\label{competingorder}

\begin{table}[htbp]
\caption{Possible momentum independent, intra-unit cell order parameters $\mathcal{O}=\langle \Psi^\dagger \hat{M} \Psi \rangle$ for a parabolic semimetal, classified according to the irreducible representations of $O_h$ point group and their properties under time-reversal symmetry (TRS) operation, where $\Psi$ is a four component spinor and $\hat{M}$ is a $4 \times 4$ Hermitian matrix that can be expressed in terms of three spin $3/2$ matrices $J_1$, $J_2$ and $J_3$. The dipolar, quadrupolar and octupolar orders are respectively abbreviated as $\mathcal{DO}$, $\mathcal{QO}$, $\mathcal{OO}$.}
\begin{center}
 \begin{tabular}{|c|c|c|}
            \hline
            Rep. & $\hat{M}$ & TRS \\
            \hline \hline
            $A_{1g}$ & $\mathbb{1}$ & $\checkmark$ \\
            \hline
            $T_{2g}$, ${\mathcal QO}$ & $\{J_1, J_2 \}, \{J_2, J_3\}, \{J_3, J_1\} $ & $\checkmark$ \\
            \hline
            $E_{g}$, ${\mathcal QO}$ & $J^2_1-J^2_2, 2 J^2_3-J^2_1-J^2_2 $ & $\checkmark$  \\
            \hline
            $A_{2u}$, ${\mathcal OO}$ & $J_1 J_2 J_3+J_3J_2J_1$ & $\times$ \\
            \hline
            $T_{1u}$, ${\mathcal DO}$  & $\left( J_1, J_2, J_3\right)$ & $\times$  \\
						\hline
            $T_{1u}$, ${\mathcal OO}$  &  $\left( J^3_1, J^3_2, J^3_3\right)$ & $\times$ \\
            \hline
            $T_{2u}$, ${\mathcal OO}$& $ J_1(J^2_2-J^2_3), J_2(J^2_3-J^2_1), J_3(J^2_1-J^2_2) $ & $\times$ \\
            \hline
    \end{tabular}
    \end{center}
    \label{table1}
    \end{table}

Next we consider the form of competing intra unit cell order parameters that can arise due to sufficiently strong short-range interactions (due to the vanishing density of states in a PSM), and their transformation properties under the time-reversal and cubic point group symmetry operations (See Table~\ref{table1}).

\subsection{Competing orders} Since the Luttinger model involves all five anticommuting $4\times 4$ $\Gamma$ matrices, we cannot write any local mass term in the particle-hole channel, which causes a uniform gap in the spectrum. This is an important difference between PSM and a Dirac semimetal (allows at least two local mass terms in the particle-hole channel). A general momentum independent, intra-unit cell order parameter can be described by $\mathcal{O}=\langle \Psi^\dagger \hat{M} \Psi \rangle$, where $\hat{M}$ is an appropriate $4 \times 4$ hermitian matrix. Any $\hat{M}$ can be expressed in terms of sixteen linearly independent matrices constructed from various combinations of spin $3/2$ matrices ($J_a$s). We can classify these linearly independent matrices into a set of six irreducible representations of the point group $O_h$, as listed in Table~\ref{table1}. Equivalently, the matrix $\hat{M}$ can be expressed in terms of five mutually anticommuting matrices ($\Gamma_j$s) appearing in the Luttinger Hamiltonian and ten product matrices defined as $\Gamma_{ij}=-i[\Gamma_i, \Gamma_j]/2$. The TRS preserving quadrupolar orders only involve even number of $J_a$s or the $\Gamma_j$s present in $\hat{H}_{L}$, and they can be grouped as a triplet and a doublet respectively following the $T_{2g}$ and $E_g$ representations ($T_{2g} \oplus E_g$). By contrast, the TRS breaking orders consist of an odd number of $J_a$s or ten $\Gamma_{ab}$ matrices, and they can follow $A_{2u}$, $T_{1u}$ and $T_{2u}$ representations ($A_{2u} \oplus T_{1u} \oplus T_{2u}$). While $A_{2u}$ and $T_{2u}$ only correspond to octupolar orders, $T_{1u}$ representation can support both dipolar and octupolar magnetic orders. Therefore, an interacting PSM can provide valuable lessons regarding competing multipolar orders and related critical phenomena observed in various strongly correlated materials.

Now we discuss this general structure of order parameters for the specific case of 227 iridates. In Appendix~\ref{sec:appendix-B}, we show that the octupolar AIAO order parameter $\varphi$ follows the $A_{2u}$ representation and couples to the low energy fermions according to 
\begin{equation}
\frac{2\varphi}{\sqrt{3}} \; \Psi^\dagger (J_1 J_2 J_3 
+ J_3 J_2 J_1) \Psi=-\varphi \; \Psi^\dagger \Gamma_{45} \Psi .\end{equation}
By contrast, the SI order parameter described by a vector $\mathbf{M}$ follows $T_{1u}$ representation, and it requires a very specific combination of dipolar and octupolar terms 
\begin{equation}
\sum_{a=1}^{3}\frac{M_a}{3} \; \Psi^\dagger \left(7J_a-4J^3_a\right)\Psi=
\sum_{a=1}^{3} M_a \Psi^\dagger \Gamma_a \Gamma_{45} \Psi.\end{equation} We also note that a general form of magnetic $T_{1u}$ order parameter gives rise to magnetic moment and can be written in terms of two vectors $\mathbf{M}_1$ and $\mathbf{M}_2$ as 
$\sum_{a=1}^{3}\Psi^\dagger[M_{1a} J_a + M_{2a} J^3_a ]\Psi$
and it naturally involves a SI component. The octupolar $T_{2u}$ order by itself cannot cause any magnetic moment. Only in the presence of additional quadrupolar order or mechanical strain, it is capable of producing a magnetic moment. However, motivated by the phenomenology of pyrochlore iridates, we will not provide any detailed discussion of quadrupolar ($T_{2g}$ and $E_g$) and octupolar $T_{2u}$ orders. 

\begin{figure*}[htbp]
\centering
\subfigure[]{
\includegraphics[width=5.75cm, height=5cm]{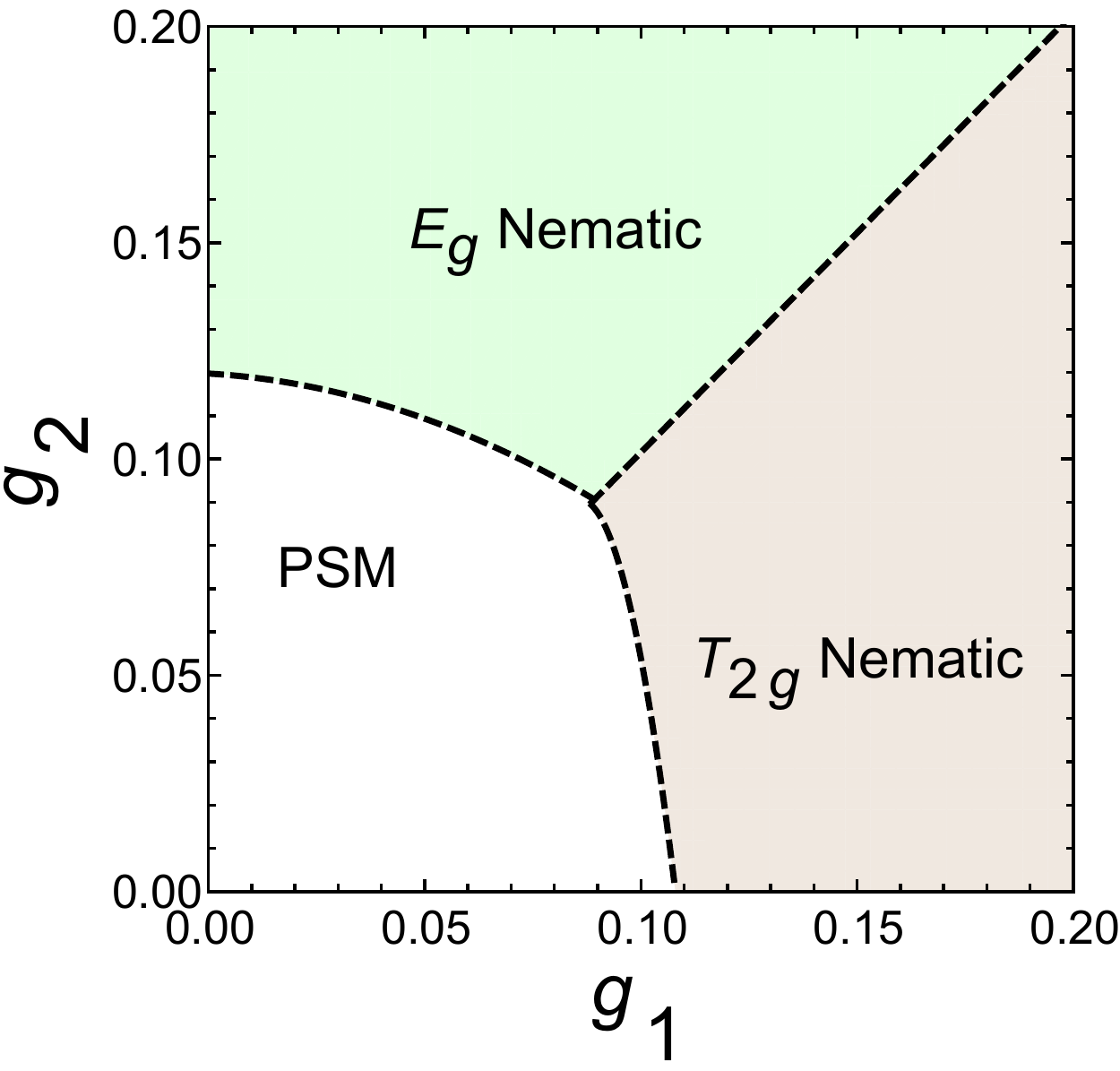}
\label{fig:compa}
}
\subfigure[]{
\includegraphics[width=5.25cm, height=5cm]{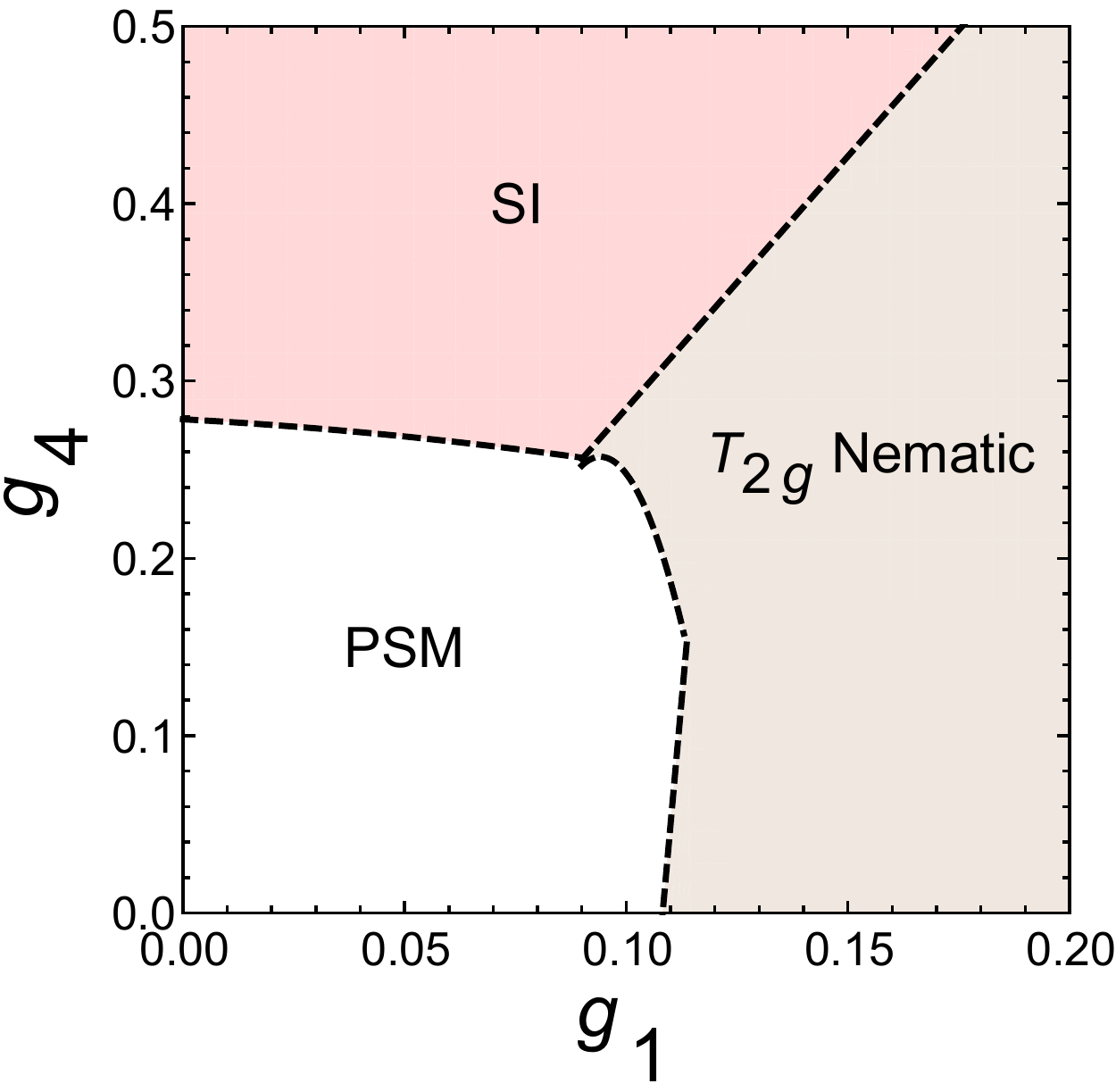}
\label{fig:compb}
}
\subfigure[]{
\includegraphics[width=5.25cm, height=5cm]{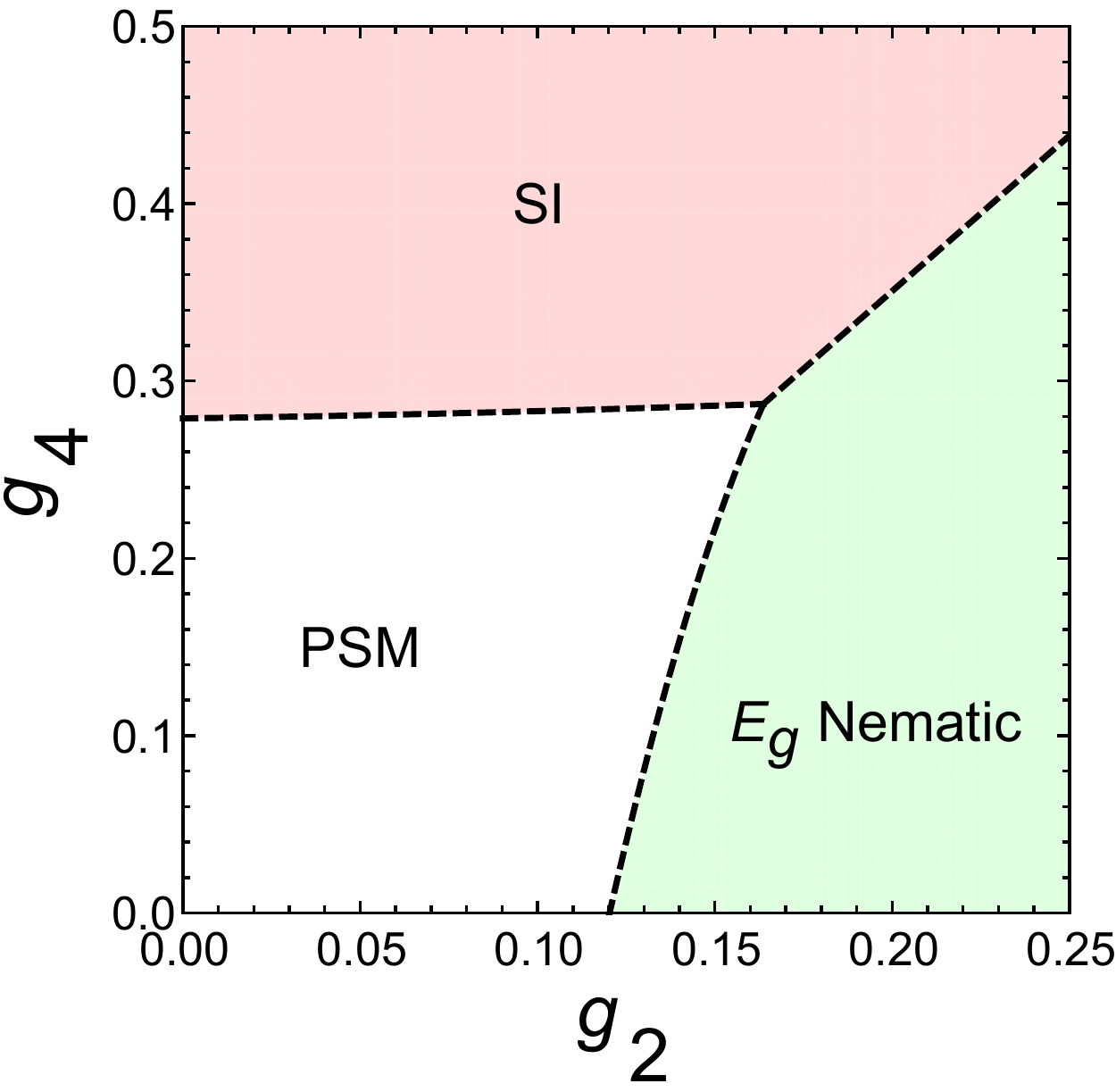}
\label{fig:compc}
}
\caption[]{ Different cuts of the global phase diagram for a strongly interacting parabolic semimetal (PSM). While extracting the phase diagram in a given two-dimensional plane of coupling constants, we have set the bare values of remaining four coupling constants to be zero. (a) The interplay between two types of nematic orders following $T_{2g}$ and $E_g$ representations of cubic point group symmetry, (b) the competition between $T_{2g}$ nematic and spin ice (SI) orders, (c) competing $E_g$ nematic and SI orders. All transitions out of the PSM phase are continuous. Even though these phase diagrams (as well as the ones shown in Fig.~\ref{fig:1e} and Fig.~\ref{Figcomp2}) have been obtained by setting $N=6$ in the renormalization group flow equations, the qualitative structure of the phase diagrams is insensitive to the specific choice of $N$.  
}\label{Fig:comp1}
\end{figure*}

\subsection{RG analysis}~\label{subsecRG}

The generic form of local density-density interactions for studying competition among different types of ordering can be described by the following model 
\begin{eqnarray}
&&H_{int}=\int d^3x \; [g_0 (\Psi^\dagger \mathbb 1 \Psi)^2 +g_1 \sum_{j=1}^{3} (\Psi^\dagger \Gamma_j \Psi)^2 \nonumber \\
&&+g_2 \sum_{j=4}^{5} (\Psi^\dagger \Gamma_j \Psi)^2 +g_3 (\Psi^\dagger \Gamma_{45} \Psi)^2 +g_4 \sum_{j=1}^{3} (\Psi^\dagger \Gamma_j \Gamma_{45} \Psi)^2 \nonumber \\ &&+ g_5 \sum_{j=1}^{3} \; \{(\Psi^\dagger \Gamma_{j4} \Psi)^2 +(\Psi^\dagger \Gamma_{j5} \Psi)^2  \}],\label{genericint}
\end{eqnarray}
where $g_\mu$s are six independent coupling constants. From a microscopic perspective, the bare value of these effective couplings are functions of onsite Coulomb repulsion $U$, nearest neighbor repulsion $V$, intra unit cell exchange interaction strengths, and also the spin orbit splitting energy of higher energy bands (as we either project or integrate out the higher energy bands). However, due to the uncertainty regarding actual microscopic model of a strongly correlated material, we will not investigate the explicit dependence of $g_\mu$s on microscopic couplings, and rather treat them as effective coupling constants, allowed by the underlying symmetry. It is natural to anticipate that sufficiently strong $g_1$ and $g_2$ drive quadrupolar orders, while large $g_3$ and $g_4$ will respectively cause nucleation of AIAO and SI order parameters.

We note that the effective action for the interacting model, defined as 
\begin{equation}
S_{int}= \int d^3 \mathbf{x} d\tau \; \Psi^\dagger \hat{H}_L \Psi + \int d\tau H_{int},
\end{equation}
remains invariant under the scale transformations 
\begin{eqnarray}
\mathbf{x} \to \mathbf{x} e^l, \; \tau \to \tau e^{z l}, \; \Psi \to \Psi e^{-dl/2}, \; g_\mu \to g_\mu e^{(d-z)l},\label{rescaling}
\end{eqnarray}
with $d=3$ and $z=2$, where $\tau$ is the imaginary time. Therefore, at the tree level all local interactions have negative scaling dimension $[g_\mu]=(z-d)=-1$, indicating the stability of a PSM against sufficiently weak, but generic form of short-range interactions. This is tied to the vanishing density of states of a PSM [recall that $\rho(E) \sim |E|^{d/z-1}=|E|^{1/2}$]. Therefore, onset of a broken symmetry phase can only occur beyond a critical strength of interactions $\sim O(\epsilon)$, where $\epsilon=(d-z)=1$. In principle, the RG analysis can be controlled by the dimensional parameter $\epsilon=(d-2)$, since interactions are marginal for a two dimensional PSM (as realized in bilayer graphene supporting instability toward ordered states for infinitesimally weak interactions~\cite{roy-classification}). For obtaining a general theoretical structure we will also consider $N$ number of fermion flavors (or quadratic band touching points), which can provide an additional control parameter for the RG calculations. This is conceptually important since $\epsilon=1$ for the physical problem of three dimensional PSM. 

We will first gain some insight into the ordering tendencies from a mean-field analysis. At the mean-field level, which corresponds to taking $N \to \infty$ limit, the instability toward broken symmetry states can be addressed from the bare susceptibilities of the order parameters. The susceptibility of AIAO and SI order parameters (at zero external frequency and momentum) are respectively given by
\begin{eqnarray}
\chi^{-1}_{3}(\mathbf{q}=0,\Omega=0)&=&\frac{\hbar^4}{8m^2_2} \; \sum_{j=4}^{5} \; \int \frac{d^3k}{(2\pi)^3} \; \frac{d^2_j(\mathbf{k})}{\epsilon^3(\mathbf{k})}, \\
\chi^{-1}_{4}(\mathbf{q}=0,\Omega=0)&=&\frac{\hbar^4}{12m^2_1} \; \sum_{j=1}^{3} \; \int \frac{d^3k}{(2\pi)^3} \; \frac{d^2_j(\mathbf{k})}{\epsilon^3(\mathbf{k})}.
\end{eqnarray}
For a rotational symmetric PSM with $m_1=m_2$, SI and AIAO channels possess equal susceptibilities
\begin{eqnarray}
\chi^{-1}_{3}(\mathbf{q}=0,\Omega=0)=\chi^{-1}_{4}(\mathbf{q}=0,\Omega=0)=\frac{4\sqrt{2} \; m_p \Lambda}{5 \hbar^2 \pi^2}, \label{AIAOSIcritical}\nonumber \\
\end{eqnarray}
where $\Lambda$ is the ultraviolet momentum cutoff, indicating why a strong competition can arise between these two channels. Even though $m_1$ and $m_2$ are not generically equal, long range Coulomb interaction tends to reduce the effects of cubic anisotropy, leading to $m_1 \sim m_2$~\cite{Abrikosov1,Abrikosov2,Balents4,Herbut1,Herbut2,lai-roy-goswami}. The linear-$\Lambda$ dependence of the susceptibilities reflects the negative scaling dimension of local four-fermion interactions ($[g_\mu]=-1$) and for the isotropic model SI and AIAO orders have equal critical couplings $g_{c3}=g_{c4}=5\sqrt{2}/64$ as shown in Fig.~\ref{fig:1d}, where we have defined dimensionless coupling constants $g_\mu m_p \Lambda/(8 \pi^2 \hbar^2) \to g_\mu$.  However in the presence of cubic anisotropy ($m_1 \neq m_2$), the degeneracy of the critical couplings is lifted. For $m_1< m_2$, the SI ordering requires a smaller critical coupling ($g_{c4}<g_{c3}$). By contrast, $m_2<m_1$ favors AIAO order as the dominant instability with $g_{c3}<g_{c4}$. We note that all magnetic bilinears $\Psi^\dagger \Gamma_{ij} \Psi$s have equal susceptibilities for the isotropic model. Similarly, for the rotational symmetric model, the susceptibilities for all quadrupolar orders are also equal. For $m_1 \neq m_2$, the susceptibilities of $T_{2g}$ and $E_g$ channels will be different from each other. `Specifically for $m_1<m_2$ the susceptibility of $T_{2g}$ nematic order is larger and consequently its nucleation requires a smaller critical coupling compared to  $E_g$ nematic order. By contrast, for $m_2<m_1$ condensation of $E_g$ nematic order requires smaller critical strength of interaction.

A better understanding of the instability of PSM toward competing broken symmetry states can be achieved through one loop RG calculations for the general interacting Hamiltonian $H=H_L+H_{int}$. The RG analysis allows us to go beyond the saddle point ($N=\infty$) or mean-field calculations and systematically account for the interplay among different ordering channels or fluctuation effects. For simplicity we will consider the isotropic model with $m_1=m_2$. We perform a Wilsonian frequency-momentum shell decimation, while manifestly preserving the $z=2$ structure of the theory. Therefore, we integrate out the fast degrees of freedom within the shell $E_c e^{-l} < \sqrt{\omega^2+(k^2/[2m])^2}<E_c$, where the high-energy cutoff $E_c \sim \Lambda^2/(2m)$. Due to the technical nature, the details of RG calculation and relevant Feynman diagrams [see Fig.~\ref{FeynmanDiagram_RG}] are shown in Appendix~\ref{sec:appendix-RG}. The RG flow equations are obtained after the subsequent rescaling of the effective action according to Eq.~(\ref{rescaling}), which have the general structure 
\begin{equation}
\frac{dg_\mu}{dl}=-\epsilon g_\mu+ N \; a_\mu \; g^2_\mu + b_{\mu \nu} g_\mu g_\nu,
\end{equation}
with $\epsilon=(d-2)=1$. The expressions for the coefficients $a_{\mu}$ and $b_{\mu \nu}$ are very lengthy and they are shown in Appendix~\ref{sec:appendix-RG}. We have also derived the RG flow equations for order parameter susceptibilities. In the flow equations, $N \; a_{\mu}$ represents the contribution coming from fermion bubble, whereas $b_{\mu \nu}$s arise from different types of vertex corrections and capture the interplay among different ordering channels. In the $N \to \infty$ limit, when all vertex correction terms can be ignored, we reproduce the mean-field results. The deviation from mean-field analysis due to fluctuations are controlled by the parameter $1/N$. From the numerical solution of flow equations, we have checked that all qualitative aspects of competing orders remain robust even for small values of $N$.

\begin{figure}[htbp]
\centering
\subfigure[]{
\includegraphics[width=4cm, height=4cm]{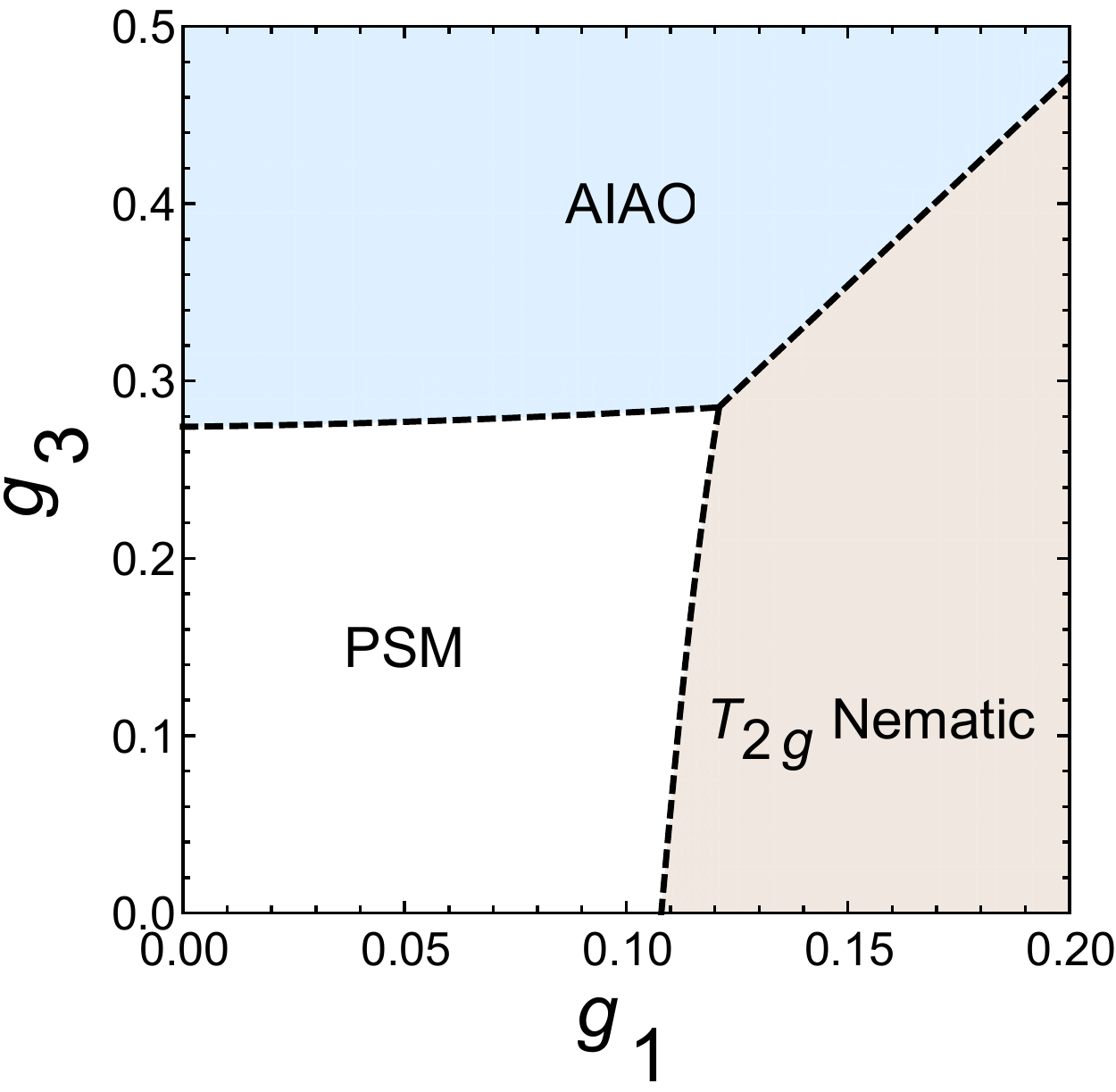}
\label{fig:compd}
}
\subfigure[]{
\includegraphics[width=4cm, height=4cm]{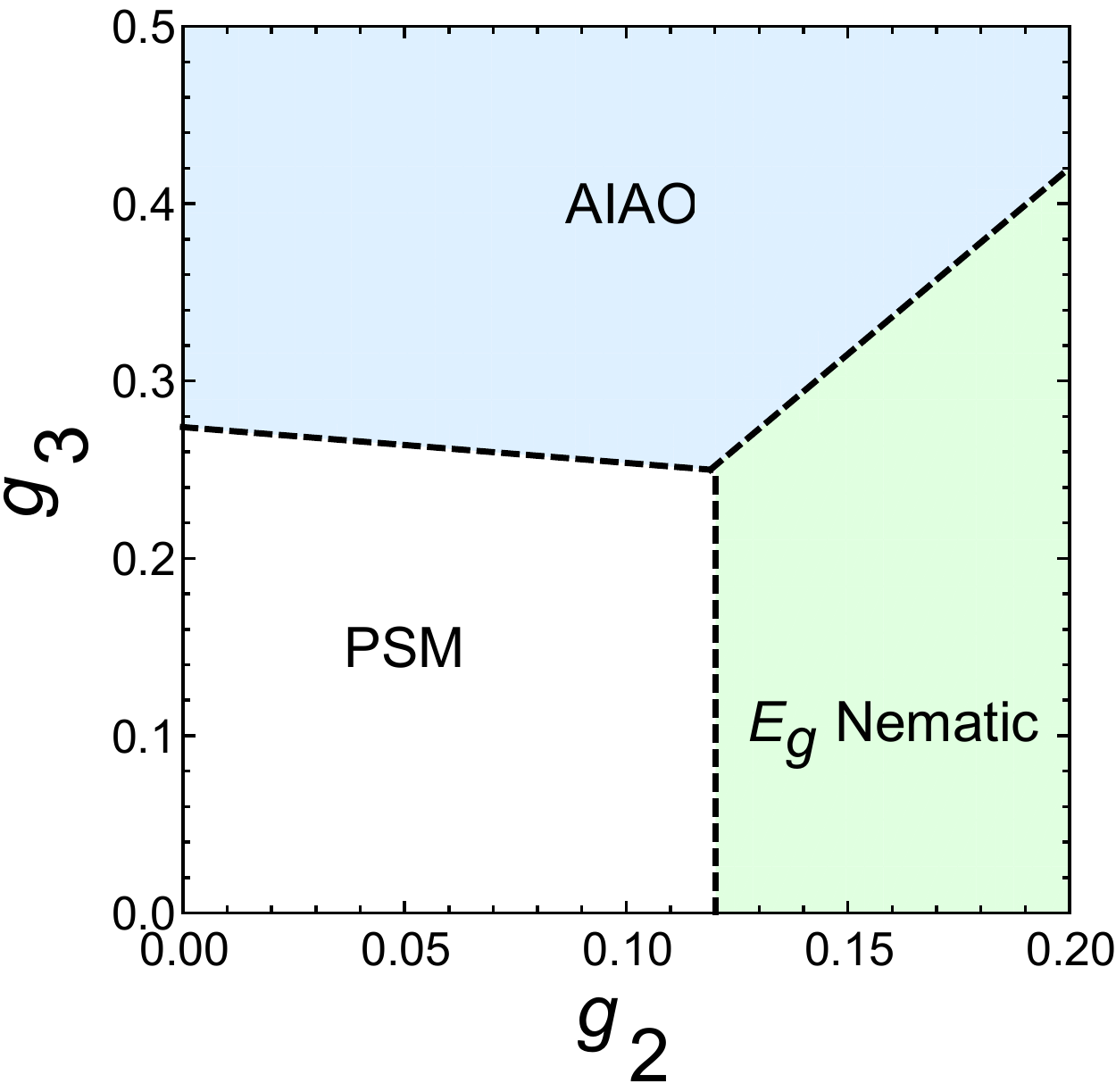}
\label{fig:compe}
}
\caption[]{ Two cuts of the global phase diagram of a three dimensional, strongly interacting parabolic semimetal (PSM) capturing the interplay of (a) $T_{2g}$ nematic and (b) $E_g$ nematic orders with all-in all-out (AIAO) order. The phase diagrams have been obtained from by solving renormalization group equations for fermion flavor number $N=6$, and by setting the bare value of remaining four coupling constants to zero.  
}\label{Figcomp2}
\end{figure}

After solving the flow equations for coupling constants ($g_\mu$) and the order parameter susceptibilities, we can obtain the global phase diagram in the six dimensional coupling constant space. The dominant diverging susceptibility determines the nature of broken symmetry phase. We have already showed the instabilities of the PSM toward AIAO and SI ordered phases in the $g_3-g_4$ plane [see Fig.~\ref{fig:1e}]. The corresponding solutions have been obtained by setting the initial values of other four couplings to zero. By going to different coupling constant planes, we can similarly capture the competition among other multipolar order parameters. For example, the competition between $T_{2g}$ and $E_g$ quadrupolar or nematic orders is shown in the $g_1-g_2$ plane in Fig.~\ref{fig:compa}. On the other hand, the competition between these two nematic orders and the SI phase (a specific combination of dipolar and octupolar orders) are respectively demonstrated in Fig.~\ref{fig:compb} and Fig.~\ref{fig:compc}. By contrast, Fig.~\ref{fig:compd} and Fig.~\ref{fig:compe} capture the competitions among the octupolar AIAO and two types of nematic orders. Thus various cuts of the six dimensional global phase diagram of a strongly interacting PSM are highlighting the possibility of realizing several multipolar orders, which can also be relevant for other correlated materials, such as half-heusler compounds. However, by keeping the phenomenology of 227 iridates in mind, in this work we will mainly focus on AIAO and SI ordered states.

This RG analysis is tailored for addressing the instabilities of PSM and it shows that onset of a broken symmetry phase occurs through a continuous QPT. Within our one loop calculations, we are finding that all the itinerant quantum critical points have the same correlation length exponent $\nu=1/\epsilon=1$. Therefore, the itinerant quantum critical points (QCPs) associated with AIAO and SI orders provide us rare examples of \emph{non-Gaussian quantum criticality} in three spatial dimensions. Similar conclusions also hold for a quadrupolar or nematic QCP. The critical property of a physical observable $(O)$ (thermodynamic or transport quantity) will generally exhibit a universal scaling form
\begin{equation}
O=\delta^{x \nu} {\mathcal F} \left( \varepsilon \delta^{-\nu z}\right),
\end{equation}
where ${\mathcal F}$ is a  universal scaling function, the exponent $x$ describes the anomalous dimension of the observable, $\delta=(g-g_c)/g_c$ is the reduced distance of the tuning parameter or the coupling constant from the QCP located at $g=g_c$, and $\varepsilon$ corresponds to either temperature or frequency ($\omega$). In general, the one loop values of critical exponents $z=2$ and $\nu=1$ will receive corrections from higher loop calculations, which are however parametrically suppressed by powers of $1/N$. Such higher order calculations are needed for demonstrating that the gapless fermionic degrees of freedom remain strongly coupled to the gapless order parameter fields at the critical point, leading to the anomalous scaling dimensions for both fields. Consequently, the residue for quasiparticle pole vanishes according to $|\delta|^{\eta \; \nu}$, when the QCP is approached from the PSM side, and we have denoted $\eta$ as the anomalous scaling dimension for quadratically touching bands. Therefore, precisely at the quantum critical point, the notion of well defined quasiparticle pole will break down, and the spectral function will display a branch cut, signifying a \emph{non-Fermi liquid}. How long range Coulomb interaction affects the nature of normal state~\cite{Abrikosov1,Abrikosov2,Balents4,Herbut1,Herbut2, lai-roy-goswami} and quantum critical phenomena~\cite{Balents3,Herbut1,Herbut2} will be discussed in a future work.

For studying the nature of SI ordered phase and transition between two ordered states at strong coupling regime, when the anchoring point of PSM phase is absent, it is more useful to work with an effective Landau free energy. However before discussing the detailed form of Landau free energy, we will consider the fermion dispersion relations within the relevant broken symmetry phases (AIAO and SI), which allows us to understand the interplay between the nodal topology and AHE, relevant for the physics of Pr$_2$Ir$_2$O$_7$.

\section{Fermion spectra and AHE}\label{AHE}

Since any magnetic ordering removes the Kramers degeneracy of the conduction and valence bands, it is capable of producing Weyl nodes (recall Fig.~\ref{fig:bande}). It is worth pointing out that a simultaneous diagonalization of five $\Gamma_j$ (appearing in the Luttinger model), three $\Gamma_{45} \Gamma_a$  (a=1,2,3) (describing the SI order) and $\Gamma_{45}$ (describing the AIAO order) matrices is identical to the digonalization of a massive Dirac Hamiltonian in the presence of combined axial scalar and vector potentials~\cite{GoswamiTewari1}. In the presence of both SI and AIAO orders, the fermion dispersion relations are obtained by solving the following quartic equation
\begin{eqnarray}
&& \left[E^2-\frac{\hbar^4}{4m^2_1}\sum_{j=1}^{3} d^2_j-\frac{\hbar^4}{4m^2_2}\left(d^2_4+d^2_5 \right)-\left(\varphi^2-\mathbf{M}^2\right)\right]^2 \nonumber \\
&+&4 \bigg[ E^2-\frac{\hbar^4}{4m^2_1}\sum_{j=1}^{3} d^2_j\bigg](\varphi^2-\mathbf{M}^2) \nonumber \\
&-& 4 \bigg[ E\varphi-\frac{\hbar^2}{2m_1}\sum_{j=1}^{3} d_j M_j\bigg]^2 =0,\label{seculargeneral}
\end{eqnarray}
where $E=E(\mathbf{k})-E_{0}(\mathbf{k})$. Now we will separately consider the fermion spectra and the nodal topology for each order. For brevity we have dropped explicit $\mathbf k$ dependence of $d_j(\mathbf k)$s.

\begin{figure}[htb]
\includegraphics[scale=0.55]{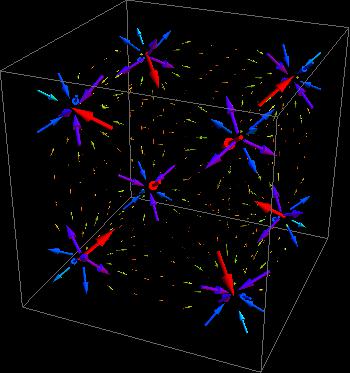}
\caption[]{Distribution of Abelian Berry curvature for a Weyl semimetal arising from an underlying ``all-in all-out" (AIAO) order. Since AIAO order preserved cubic symmetry, four right handed and four left handed Weyl points are symmetrically located on the diagonals of a cube. Right and left handed Weyl points respectively act as the source and sink of Berry curvature. Due to cubic symmetry, the net Berry flux through any plane vanishes, and a pure AIAO ordered state cannot support anomalous Hall effect.}\label{berry1}
\end{figure}

\subsection{AIAO order}

The pure AIAO order ($\mathbf{M}=0$ and $\varphi \neq 0$) modifies the fermion spectra according to
\begin{eqnarray}
&&E_{n,s}(\mathbf{k})=E_0(\mathbf{k}) + \mathrm{sgn}(n) \; \frac{\hbar^2}{2 m_p} \: \bigg [\sin^2 \alpha \; (d^2_4+d^2_5) \nonumber \\ &+&\left[\cos \alpha \; \sqrt{\sum_{j=1}^{3}d^2_j} + s \; \left(\frac{2 m_p \varphi}{\hbar^2}\right)\right]^2\bigg]^{1/2}.
\end{eqnarray}
The conduction and valence bands denoted by $n=\pm 1$ and $s=-1$ touch at eight Weyl points located at $k_0(\pm 1, \pm 1, \pm 1)$, where $k_0= \sqrt{2 m_1 \varphi/(3\hbar^2)}$ [as illustrated in Fig.~\ref{berry1}]. Therefore, the low energy thermodynamic and transport properties of the AIAO phase are governed by the linearly dispersing Weyl fermions. However, the net Berry flux through any plane vanishes due to the preserved cubic symmetry. Consequently, the AIAO order does not support any AHE.

\begin{figure}[htb]
\includegraphics[scale=0.35]{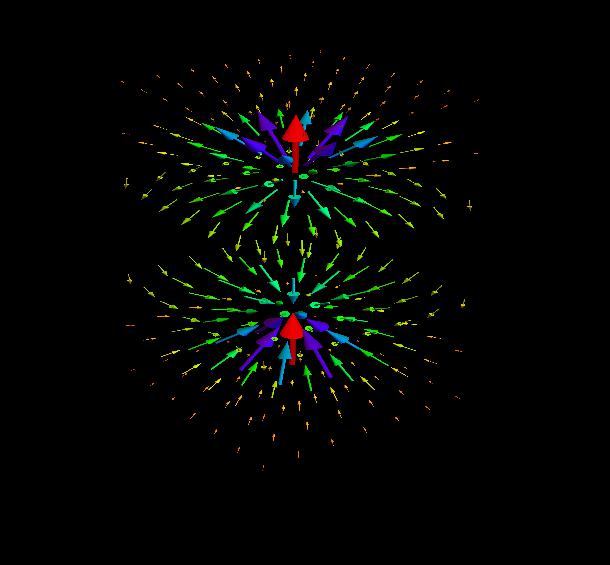}
\caption[]{ Distribution of Abelian Berry curvature for a Weyl semimetal arising in the presence of an underlying uniaxial spin ice (SI) order. Due to the broken cubic and time reversal symmetries, the Berry flux through a plane perpendicular to the direction of nodal separation is quantized to be $2\pi$, and the SI ordered phase can support a large anomalous Hall effect. This also holds for ``three-in one-out" phase obtained by locking the vector order parameter of SI state along one of the eight possible [1,1,1] directions. The Weyl points are then separated along the [1,1,1] direction, giving rise to a large anomalous Hall effect in the perpendicular plane, as observed in Pr$_2$Ir$_2$O$_7$. }\label{berry2}
\end{figure}

\subsection{SI order} The uniaxial SI order gives rise to a nodal ring in addition to two Weyl nodes. For example, the uniaxial SI order with $\mathbf{M}=(0,0,M_3)$ leads to the spectra
\begin{eqnarray}
&&E_{n,s}(\mathbf{k})=E_0(\mathbf{k}) + \mathrm{sgn}(n) \; \frac{\hbar^2}{2m_p} \; \bigg [\cos^2 \alpha \; (d^2_1+d^2_2)  \nonumber \\ &+&\left[\sqrt{\cos^2 \alpha \; d^2_3+\sin^2 \alpha (d^2_4+d^2_5)}+ s \; \left(\frac{2 m_p M_3}{\hbar^2}\right)\right]^2\bigg]^{1/2}. \nonumber \\
\end{eqnarray}
Consequently, two bands with $n= \pm1$, $s=-1$ touch at two Weyl points located at  $(0,0,\pm \sqrt{2 m_2 M_3}/\hbar)$, and also along a line node in the $k_3=0$ plane.  Similar coexistence of two different types of nodal quasiparticles have been discussed in the context of $k_z(k_x \pm ik_y)$ and $k_z(k_x \pm ik_y)^2$ pairings for URu$_2$Si$_2$~\cite{GoswamiBalicas} and UPt$_3$~\cite{GoswamiAndriy} respectively. The low energy density of states for the line and the Weyl nodes respectively behave as $\rho(E) \sim |E|$ and $\rho(E) \sim E^2$. Consequently, the thermodynamic properties for the uniaxial SI ordered phase at low temperatures are dominated by the line node. However, due to the absence of cubic symmetry inside SI phase, the Weyl nodes produce a quantized $2\pi$ Berry flux through the $xy$ plane. The distribution of Berry curvatures for uniaxial SI phase is shown in Fig.~\ref{berry2}, which causes an anomalous Hall current~\cite{GoswamiTewari1,Ran,Burkov}
\begin{equation}
\mathbf{j}_H=\left(\frac{e^2}{h}\right) \; \frac{\Delta \mathbf{k}}{2\pi} \times \mathbf{E}, \label{ahesigma}
\end{equation}
where $\mathbf{E}$ is the external electric field and $\Delta \mathbf{k} \parallel \mathbf{M}$ is the nodal separation vector pointing from the left handed Weyl point to the right handed Weyl point. Therefore the direction of AHE is determined by the direction of the net magnetic moment. For the uniaxial SI order with $\mathbf{M}=(0,0,M_3)$ we obtain the anomalous Hall conductivity
\begin{equation}
\sigma_{xy}= -\sigma_{yx}=- \; \mathrm{sgn}(M_3) \; \frac{2 e^2}{h^2} \; \sqrt{2 m_2 M_3}. \label{HallSI}
\end{equation}
\emph{Notice that the anomalous Hall conductivity depends on the square root of the magnetization or the magnetic moment}. Therefore a small order parameter or magnetic moment can cause an order of magnitude enhancement of anomalous Hall conductivity in contrast to the expectations of conventional Karplus-Luttinger theory that $\sigma_H$ varies linearly with the magnetic moment. The Weyl fermions also give rise to anomalous thermal Hall and anomalous Nernst effects in the $xy$ plane, in addition to the polar Kerr effect and Faraday rotation in optical measurements~\cite{GoswamiTewari1,GoswamiTewari2}. However a uniaxial SI order can only be nucleated by applying a sufficiently strong magnetic field along one of the C$_{4v}$ axes, as discussed in Subsection~\ref{magfield}. However, for such a field induced state the notion of AHE is not meaningful. But it still leaves the option of probing power law behaviors of thermodynamic and transport properties due to the line node (provided the strong field does not cause sharp Landau levels). 

However, the observed AHE along [1,1,1] direction in Pr$_2$Ir$_2$O$_7$ cannot be explained in terms of a uniaxial SI order. Later based on the Landau theory we will show that the underlying cubic environment locks the SI order parameter vector $\mathbf{M}$ along one of the eight possible $[1,1,1]$ directions. We can call such a SI ordered state as triplet spin ice (TSI), which is a particular form of 3I1O order. After locking $\mathbf{M}$ along [1,1,1] direction, the system decreases the number of gapless quasiparticles by removing the line node, leading to a gain in condensation energy. This can be seen from the dispersion relations of 3I1O phase.

\subsection{3I1O order}

We first consider the simplest form of 3I1O order that corresponds to $\varphi=0$, $\mathbf{M}=M[1,1,1]/\sqrt{3}$ ($[1,1,1]$ TSI order). Then the fermion spectra are given by
\begin{eqnarray}
&& E_{n,s}(\mathbf{k})=E_0(\mathbf{k}) + \mathrm{sgn}(n) \; \frac{\hbar^2}{2 m_p} \; \bigg[ \cos^2 \alpha \; \sum_{j=1}^{3}d^2_j \nonumber \\
&+& \sin^2\alpha \; (d^2_4+d^2_5) +  \left(\frac{2 m_p M}{\hbar}\right)^2 +2 s  \left(\frac{2 m_p M}{\hbar} \right) \nonumber \\
&\times& \sqrt{\sin^2 \alpha \; (d^2_4+d^2_5) +\frac{\cos^2 \alpha}{3} \; \left(\sum_{j=1}^{3}d_j\right)^2} \; \bigg]^{1/2}.
\end{eqnarray}
Two bands with $n= \pm1$, $s=-1$ touch only at a pair of Weyl points located at $\pm (k_0,k_0,k_0)$, with $k_0= \sqrt{2 m_1 M/(3\hbar^2)}$. As TSI order breaks both cubic and time reversal symmetries, it can support a large AHE. By applying Eq.~(\ref{ahesigma}) we find the Hall conductivity is determined by the tensor
\begin{eqnarray}
\sigma_{ab}=-\frac{2e^2}{h^2} \; \sqrt{2m_1 \; M} \; \epsilon_{abc} \; \hat{M}_c.
\end{eqnarray}  
For $\hat{M}=[1,1,1]/\sqrt{3}$ we find
\begin{eqnarray}
\sigma_{xy}=\sigma_{yz}=\sigma_{zx}=-\frac{2e^2}{h^2} \; \sqrt{\frac{2m_1 \; M}{3}},\label{AH111}
\end{eqnarray} 
and the maximum value of Hall conductivity occurring in a plane perpendicular to [1,1,1] direction is determined by
\begin{equation}
\sigma_{H}= \frac{1}{\sqrt{3}} \; (\sigma_{xy}+\sigma_{yz}+\sigma_{zx})= -\frac{2 e^2}{h^2} \; \sqrt{2m_1 M}. \label{AHE111}
\end{equation}
Notice that the direction of AHE is fully consistent with the direction of underlying magnetic moment. \emph{We note that the noncoplanar spin arrangement for these magnetic orders can be used to define a scalar chirality for each of the four faces of the tetrahedron.  and net chirality. The net chirality has the same direction as that of magnetic moment or the direction of Berry flux.}

In the presence of both TSI and AIAO orders, the $E$-linear term in the secular equation [see Eq.~(\ref{seculargeneral})] leads to cumbersome analytical expressions for four bands. When $|\mathbf{M}|>> |\varphi|$, the small AIAO component tends to shift the reference energies of two Weyl nodes. Therefore the right and the left handed Weyl fermions respectively produce electron and hole pockets even at half-filling. If the amplitude of AIAO order is much bigger than the TSI amplitude $|\mathbf{M}|$, the net Berry flux through the plane perpendicular to [1,1,1] direction becomes nonzero. However, the resulting anomalous Hall conductivity is much smaller than the one found for pure TSI phase. For conventional 3I1O order (corresponding to $|\varphi|=M$) we find simpler expressions
\begin{eqnarray}
&&E_{n,s}(\mathbf{k})=E_0(\mathbf{k}) - s \; \mathrm{sgn}(\varphi) \;  M  + \mathrm{sgn}(n) \; \frac{\hbar^2}{2 m_p} \bigg \{\sin^2 \alpha \nonumber \\ &\times &(d^2_4+d^2_5)+\cos^2 \alpha \; \sum^{3}_{j=1} d^2_j+\left(\frac{2 m_p M}{\hbar}\right)^2 \nonumber \\ &+& s \; \frac{2m_p \; M}{\sqrt{3} \hbar^2} \cos \alpha \sum_{j=1}^{3} d_j \bigg \}^{1/2}
\end{eqnarray}
for the spectra, from which we can clearly see that the two Weyl points along [1,1,1] direction are shifted in energy by an amount $\Delta E= 2 \varphi = 2 \mathrm{sgn}(\varphi) \;  M$. Similar energy shift also occurs for the other six Weyl points discussed in the context of AIAO order. Finally, the particle-hole anisotropy term makes the volume of electron and hole pockets different.

We also note that any general form of $T_{1u}$ order is capable of producing AHE. For example, one can consider a purely dipolar ferromagnetic order, which couples to the parabolic fermions according to $\mathbf{N} \cdot \Psi^\dagger \mathbf{J} \Psi$. The underlying cubic environment will also lock $\mathbf{N}$ along one of the [1,1,1] directions. However, the observed SI correlations cannot be explained in terms of a pure dipolar ferromagnetic order, and this is one of our main reasons for considering SI and 3I1O orders. 
Now we discuss how 3I1O order can account for the observed AHE in Pr$_2$Ir$_2$O$_7$.

\section{Physics of $\mathrm{Pr}_2\mathrm{Ir}_2\mathrm{O}_7$}\label{section-pr}

Based on the discussion in previous section we infer that any itinerant SI order can give rise to a large AHE. The amplitude of the SI order parameter ($|\mathbf{M}|$) at $T=0$ should be roughly given by $k_B T_H$ where $T_H$ is the onset temperature for the AHE, which can be estimated in the following manner. The experimentally observed anomalous Hall conductivity defines a length scale
\begin{equation}
l_H=\frac{e^2}{h} \times \frac{1}{\sigma_{H}} \approx 3.8 \times 10^{-8} m.
\end{equation}
Since the PSM displays $z=2$ scaling between length and time scales, we can equate $l_H$ with the thermal de Broglie wavelength of the PSM at $T_H$. This leads us to 
\begin{equation}
|\mathbf{M}|\approx k_B T_H=\frac{h^2}{2m^\ast \; l^2_H}  =\frac{h^4 \; \sigma^2_H}{2m^\ast \; e^4}, \label{tempHall}
\end{equation}
where $m^\ast$ is the effective mass of the quadratically touching bands. \emph{Just based on appropriate scaling analysis and without any explicit calculations we find} 
$$\sigma_H \approx \frac{e^2}{h^2} \; \sqrt{2 m^\ast \; |\mathbf{M}|},$$
\emph{which almost quantitatively matches the anomalous Hall conductivities of the Weyl semimetals obtained for SI and 3I1O ordered states in the previous section.} The experimentally observed large $\sigma_H$ and smallness of moment $|\mathbf{m}|$ can be easily reconciled by invoking itinerant SI order. However, the direction of the AHE can only be understood in terms of itinerant 3I1O order.


Even though a general form of 3I1O order along [1,1,1] direction will be consistent with the experimentally observed direction of AHE, for extracting qualitative estimates of the associated energy scales we will consider the simplest form of 3I1O order, as described by the [1,1,1] TSI configuration. If experimentally determined Hall conductivity $10^3 \; \Omega^{-1} m^{-1}$ is associated with $\sigma_{xy}=\sigma_{yz}=\sigma_{zx}$ of Eq.~(\ref{AH111}) the amplitude for TSI order parameter will be
$$| \mathbf M| \approx 7.2 \times 10^{-31}/m_1 \; \mathrm{meV}. $$
Recent ARPES experiment does not find any evidence of strong cubic anisotropy, suggesting $m_1 \sim m_2 \approx 6.3 \; m_e$. By choosing this value of $m_1$ we will obtain
$$|\mathbf M| \approx 0.13 \; \mathrm{meV} \: \Rightarrow T_H \approx 1.5 \mathrm{K},$$
which is consistent with the temperature scale for the onset of an AHE, as observed in Pr$_2$Ir$_2$O$_7$~\cite{Nakatsuji2, Nakatsuji3, Nakatsuji4}. If we rather take the maximum value of anomalous Hall conductivity $\sigma_H$ of Eq.~(\ref{AHE111}) to be $10^3 \; \Omega^{-1} m^{-1}$, the amplitude of the TSI order will be three times smaller and $T_H \sim 0.5 K$. The order parameter amplitude, and the magnetic moment $\mathbf{m}$ per Ir ion are roughly related as
\begin{eqnarray}
\mathbf{m} \approx \frac{ g_L \; \mu_B}{g_4} \; \mathbf{M} \label{moment1}
\end{eqnarray}
where $g_L \sim 2$ is the Lande $g$ factor for 5d electrons, $\mu_B$ is the Bohr magneton, and $g_4$ is the effective strength of interaction in the SI channel. An interaction energy scale $g_4 \sim 1$ eV, produces an estimated magnetic moment $\sim 10^{-5} \mu_B$per Ir ion. Since experimental measurement of magnetization accounts for the net moment coming from both Ir and Pr sublattices, we now consider the role of Pr$^{3+}$ ions.

The small magnetic moment of Ir will also produce a small effective field along [1,1,1] direction for Pr ions through the underlying Kondo coupling. If the Kondo coupling is antiferromagnetic, magnetic moments of Ir and Pr will tend to align along opposite directions, which can cause further suppression of the net magnetic moment of the system. Our estimated $\mathbf{m}$ is therefore consistent with experimental bound~\cite{Nakatsuji2, Nakatsuji3, Nakatsuji4} ($<10^{-3} \mu_B$). When an external magnetic field is applied along [1,1,1] direction, Pr moments will tend to point along the external field due to a larger Zeeman coupling, while Ir moment will align along [-1,-1,-1]. Therefore, the sign of AHE can be negative. At least this occurs for a large value of external magnetic field~\cite{Nakatsuji4}. Inside the TSI phase, the induced magnetic moment on Pr is nonuniversal and can be quite small due to several reasons. The product of small itinerant Ir moment and a small scale of Kondo coupling produces a very tiny effective field for Pr ions along [1,1,1] direction, and it is inadequate to order Pr local moments. This will certainly be compatible with antifferromagnetic exchange coupling between local moments as suggested by the negative $T_{CW}$. The moments of Pr ion can also be suppressed by geometric factors related to the unit cell and the actual details of the Pr ions. The nonuniversality of magnetic moment for the lanthanide ions have also been observed inside the AIAO ordered insulating state. For example, $T_{MI}$ for Tb$_2$Ir$_2$O$_7$ and Er$_2$Ir$_2$O$_7$ are respectively 130 K and 140 K, but only the Tb$^{3+}$ exhibits a magnetic moment below $T \sim 40 K$~\cite{Chapon}. Therefore we believe that the most important aspects of Pr$_2$Ir$_2$O$_7$ in the absence of external magnetic field arise from Ir sector. Only for a sufficiently strong external magnetic field the Pr ions will play any important role. Now we consider the effects of competing SI and AIAO orders on the global phase diagram of 227 pyrochlore iridates within the framework of Landau theory.

\section{Landau theory of competing orders}\label{landautheory}

\begin{figure}[htb]
\includegraphics[scale=0.2]{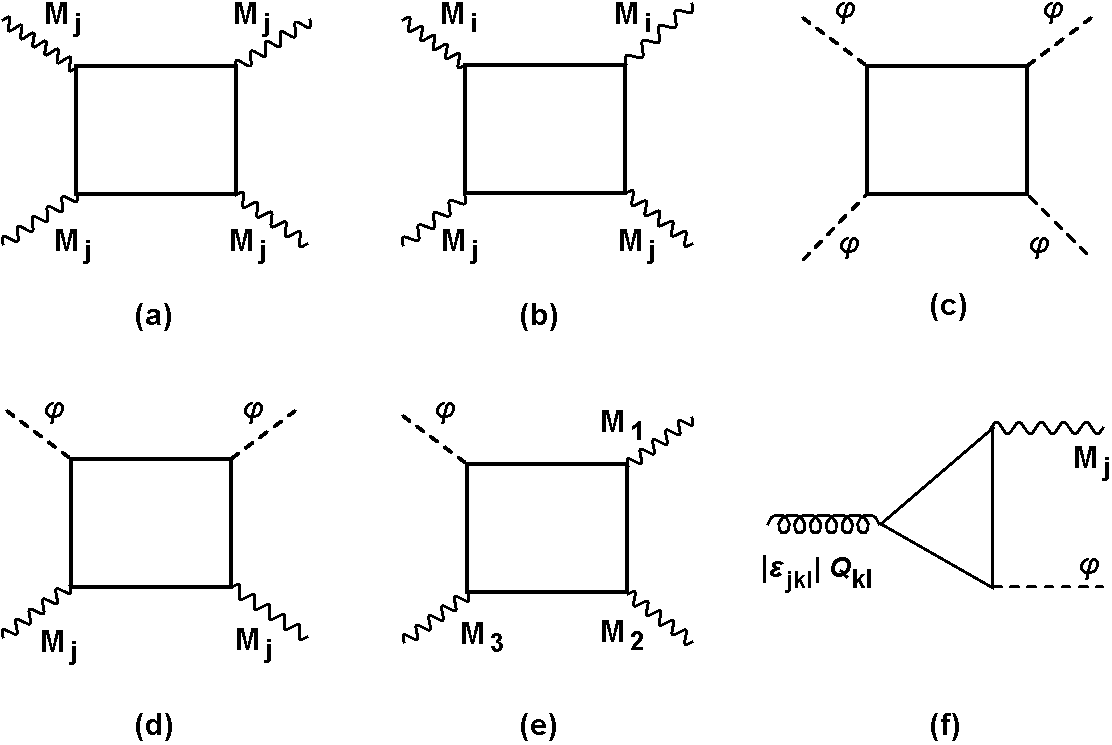}
\caption[]{ Feynmann diagrams in (a), (b), (c), (d) and (e) respectively give rise $u^\prime_1 M^4_j$, $u^{\prime \prime}_{1} M^2_i M^2_j$, $u_2 \varphi^4$, $u_{12} \varphi^2 M^2$ and $u^{\prime}_{12} \varphi M_1 M_2 M_3$ terms. The triangle diagram in (f) gives rise to the coupling among the $T_{2g}$ quadrupolar, SI and AIAO orders. }\label{diag}
\end{figure}

The form of the Landau free energy and the resulting phase diagram crucially depend on the presence or absence of particle hole symmetry in the normal state. Therefore we consider these two situations separately in the following two subsections.

\subsection{Particle-hole symmetric semimetal} 

For simplicity we begin by considering the particle-hole symmetric situation, when the band structure is entirely determined by $d_j(\mathbf{k})$ and $E_0(\mathbf{k})=0$. When the Fermi level is exactly pinned at the band-touching point, the SI and AIAO orders can only couple through $\varphi^2 \mathbf{M}^2$ term at the quartic order. The explicit form of the free energy density in the absence of an external magnetic field and mechanical strain is given by
\begin{eqnarray}
f&=& r_1 \mathbf{M}^2 +3 u^{\prime}_{1}\sum^{3}_{j=1} M^4_j +  3 u^{\prime \prime}_{1}\sum_{i<j} M^2_i M^2_j \nonumber \\ & + &r_2 \varphi^2 +u_2 \varphi^4+2u_{12} \mathbf{M}^2 \varphi^2,\label{lfph}
\end{eqnarray}
where $r_1=r_{01}(T-T_{c,1})$, $r_2=r_{02}(T-T_{c,2})$, and $r_{0j}$s are positive constants. The transition temperatures of the SI and AIAO orders are respectively denoted by $T_{c,1}$ and $T_{c,2}$. The positive constants $u^{\prime}_{1}$, $u^{\prime \prime}_{1}$, $u_2$ and $u_{12}$ are functions of interaction strength and band parameters, which are respectively obtained from the Feynman diagrams (a), (b), (c) and (d) in Fig.~\ref{diag}. Due to the cubic anisotropy $u^{\prime}_{1} \neq 2 u^{\prime \prime}_1$. For orientation recall the phase diagram of Fig.~\ref{fig:1e} obtained from RG analysis of interacting PSM. In the presence of weak $g_3$ (smaller than critical strength for the onset of AIAO) and large $g_4$ (bigger than critical strength for the onset of SI) the system will be in the SI phase, corresponding to $r_1<0$ and $r_2>0$. Similarly, a strong $g_3$ and weak $g_4$ lead to $r_1>0$ and $r_2<0$, implying the AIAO phase. By contrast, when both $g_3$ and $g_4$ are large, PSM cannot be present as a zero temperature phase, and both $r_1$ and $r_2$ will possess negative values.

We first consider the parameter regime $r_1<0$ but $r_2>0$ that only supports the SI order. In the SI phase, the cubic anisotropy locks the vector order parameter $\mathbf{M}$ along one of the eight possible $C_{3v}$ directions $[\pm 1, \pm 1, \pm 1]$ (since $u^{\prime}_{1} \neq 2 u^{\prime \prime}_1$). Consequently, the ordered state does not support Goldstone modes, and it is rather described by an O(3) order parameter with $Z_8$ anisotropy. The low temperature ground state exhibits TSI order with $\mathbf{M}=(M/\sqrt{3}) \;  [\pm 1, \pm 1, \pm 1]$, where
$$M=\sqrt{-\frac{r_1}{2u_1}}, \; \mathrm{and} \; u_1=u^\prime_1+u^{\prime \prime}_{1}.$$
Due to the eight-fold degeneracy, even within the TSI phase there can be strong fluctuations of the order parameter, and one may not immediately observe a true long range order due to domain formation. By contrast, a pure AIAO order is stabilized when $r_2<0$ but $r_1>0$, with $\varphi=\sqrt{-r_2/(2u_2)}$.

The nature of the transition between two ordered states crucially depends on the relative strengths of $u_1u_2$ and $u^2_{12}$. When $u_1 u_2 > u^{2}_{12}$, there is a \emph{tetracritical} point at $T=T_{c,1}=T_{c,2}$, where four lines of second order transitions meet. Apart from the pure SI and AIAO ordered phases, there will be a coexisting phase with both $\varphi \neq 0$ and $M \neq 0$, which corresponds to a generic form 3I1O ordered state. 
On the other hand, for $u_1 u_2 <u^{2}_{12}$ the SI and AIAO orders cannot coexist, and at low temperatures they will be separated by a first order transition.  
Now a \emph{bicritical} point 
will be located at $T=T_{c,1}=T_{c,2}$, where two lines of second order transition and a line of first order transition meet. 
Next we consider the effects of the particle-hole anisotropy ($E_0(\mathbf k) \neq 0$) and an underlying Fermi surface ($\mu \neq 0$) on the phase diagram.

\subsection{Particle-hole asymmetric semimetal}

Generically the normal state possesses particle-hole anisotropy due to $E_0(\mathbf{k}) \neq 0$ and an underlying Fermi surface, which have important effects on the phase diagram when the system tends to exhibit SI order. The SI and AIAO orders can now couple through an additional quartic term
$$3 \sqrt{3} \; u^\prime_{12} \; M_1M_2M_3 \;  \varphi $$
which arises from the Feynmann diagram (e) in Fig.~\ref{diag}. We find
\begin{eqnarray}
u^\prime_{12} \propto \frac{1}{\beta} \sum_n \; \int d^3 k \; \frac{(k_1k_2k_3)^2 \; [i \omega_n +\mu-E_0(\mathbf{k})]}{[\{i \omega_n +\mu-E_0(\mathbf{k})\}^2-\epsilon^2(\mathbf{k})]^4}, \label{u12prime}
\end{eqnarray}
where $\mu$ is the underlying chemical potential, $\omega_n=(2n+1) \pi /\beta$ is the fermionic Matsubara frequency, and $\beta=1/(k_B T)$. This formula for $u^\prime_{12}$ is valid in the immediate vicinity of $\mathrm{max}\{T_{c,1},T_{c,2}\}$ inside the ordered phase.
Thus the free energy density gets modified according to
\begin{equation}
f=r_1 M^2+u_1 M^4+r_2 \varphi^2 +u_2 \varphi^4+2u_{12} M^2 \varphi^2+ u^\prime_{12} \; s \;  \varphi \; M^3 ,\label{freegen}
\end{equation}
where $s=\mathrm{sgn}(M_1M_2M_3)$. Notice that $s u^{\prime}_{12} M^3$ acts as an external field for $\varphi$. Therefore, the phase diagram can support only two phases: (i) a general 3I1O order comprised of coexisting TSI and AIAO and (ii) a pure AIAO order, as shown in Fig.~\ref{Fig2}.

When $u_{12}^{\prime} \neq 0$, for minimizing $f$ we have to solve the following set of coupled equations
\begin{eqnarray}
\varphi^3 + \left(\frac{r_2}{2u_2}+\frac{u_{12} M^2}{u_2}\right) \varphi + \frac{u_{12}^{\prime} s M^3}{4u_2}=0,\label{AIAO}\\
M\bigg[M^2 + \frac{3 s u_{12}^{\prime} \varphi }{4u_1} M +\frac{r_1}{2u_1}+ \frac{u_{12} \varphi^2}{u_1} \bigg]=0.\label{SI}
\end{eqnarray}
Notice that $M=0$ is always a solution of these equations, which can describe either the normal phase with $\varphi=0$ or a pure AIAO phase with $\varphi= \pm \sqrt{-r_2/(2u_2)}$. The pure AIAO phase is stable when $r_2$ is sufficiently negative and satisfies $|r_2|> u_2 |r_1|/u_{12}$. Now we show why the pure TSI ordered phase (when $r_1<0$ and $r_2>0$ or $T_{c,2}<T<T_{c,1}$) discussed earlier gets eliminated in the presence of $u^{\prime}_{12} \neq 0$. In this regime but just below $T_{c,1}$, we can ignore $\varphi^3$ and $\frac{u_{12} M^2}{u_2} \varphi$ terms in Eq.~(\ref{AIAO}) and $\varphi$ dependence in Eq.~(\ref{SI}). Consequently we obtain
\begin{eqnarray}
M \sim \sqrt{-\frac{r_1}{2u_1}}, \: \varphi \sim - \frac{s \; u_{12}^{\prime}}{2 r_2} \; \left[-\frac{r_1}{2u_1}\right]^{3/2}.\label{inducedAIAO}
\end{eqnarray}
Therefore, in the parameter regime $r_1<0$ and $r_2>0$, a small amplitude of TSI (just below $T_{c,1}$) induces a much smaller AIAO component, and the pure TSI phase gets converted into a generic 3I1O state. When $r_2$ is gradually tuned toward a negative value, the strength of $\varphi$ increases, while $M$ systematically decreases. When $u_1 u_2 >u^{2}_{12}$ there is a second order transition between 3I1O and pure AIAO ordered phases, and three lines of continuous transitions meet at a \emph{multicritical} point (no longer a tetracritical point), as shown in Fig.~\ref{fig:2d}. However for $u_1 u_2 <u^{2}_{12}$ the 3I1O state is separated from the AIAO phase by a first order transition at low temperatures, see Fig.~\ref{fig:2e}.

Externally applied strain and magnetic field are two important experimental tuning parameters for probing the phase diagram, shown in Fig.~\ref{Fig2}. However, these probes explicitly break the underlying cubic symmetry and can cause coupling between different sets of multipolar order parameters, discussed in Sec.~\ref{competingorder}. Therefore, an understanding of the ground state in the presence of these perturbations are extremely important for properly interpreting the corresponding experimental results. In the following section we show how Landau theory provides us valuable insight into this complex problem.

\section{Experimental consequence: effects of strain and magnetic field}\label{strain-magfield}

In this section we will elucidate the multipolar nature of the SI (superposition of $T_{1u}$ dipolar and octupolar objects) and AIAO ($A_{2u}$ octupolar quantity) orders, and their interplay with time reversal symmetric nematic or quadrupolar orders in the presence of applied external strain and magnetic fields. We consider their effects separately in the following two subsections. Additional experimental aspects are discussed in the third subsection.

\subsection{External strain}

We have emphasized that the particle-hole anisotropy leads to a generic form of 3I1O order, which is a superposition of the TSI and AIAO orders. On the general symmetry ground the coexistence of these two orders can produce a weak lattice distortion (nematicity) along [1,1,1] direction, which can be seen in the following way. The SI and AIAO orders can couple to the $T_{2g}$ quadrupolar order parameter ($Q_{ab}$) according to
\begin{equation}
f^\prime=v |\epsilon_{abc}| Q_{ab} M_c \varphi, \:\: \mathrm{where} \:\; Q_{ab}=|\epsilon_{abc}|\langle \Psi^\dagger \Gamma_c \Psi \rangle.\label{strain1}
\end{equation}
The constant $v$ can be determined by evaluating the Feynmann diagram (f) in Fig.~\ref{diag}, and only in the presence of particle-hole anisotropy it can be nonzero. For simplicity if we assume $E_{0}(\mathbf{k}) =0$, $m_1=m_2=m$ but $\mu \neq 0$ we obtain a simplified expression
\begin{eqnarray}
v &\propto& \beta \sum_{n, s} \; \int d^3k \; n \; \mathrm{sech}^2 \left[\frac{\beta}{2} \{\mu-E_{n,s}(\mathbf{k}) \}\right]  \nonumber \\
&\times& \bigg( \frac{4 m}{|\mathbf{d}|} + \beta \;  \mathrm{sech}^2 \left[\frac{\beta}{2} \{\mu-E_{n,s}(\mathbf{k}) \} \right] \bigg),\label{strain2}
\end{eqnarray}
which after performing the momentum integral yields
$$ v \sim \frac{m^{3/2} \mu }{T^{3/2}}.$$
Since both $\varphi$ and $\mathbf{M}$ are nonzero for a generic 3I1O order, their product (as encoded in $f^\prime$) acts as an external field for the $T_{2g}$ quadrupolar order. This in turn produces an expectation value for $Q_{ab}$, causing a mechanical strain of the form
$$e_{xy}=e_{yz}=e_{zx} \neq 0.$$

In the parameter regime $r_1<0$ and $r_2>0$, but just below the transition temperature $T_{c,1}$, the induced $Q_{ab}$ will be very small and $\propto M^4$. However, an externally applied $T_{2g}$ strain (along [1,1,1] direction) couples linearly to the $T_{2g}$ order parameter, causing a static expectation value for $Q_{ab}$. The induced $T_{2g}$ order parameter gives rise to a direct quadratic coupling between TSI and AIAO orders (in contrast to the quartic coupling $\propto u^\prime_{12}$). \emph{Consequently, an external strain can amplify the 3I1O order, making it possible to observe the AHE at higher temperatures.} We can estimate the enhancement of the transition temperature of 3I1O order in the following way. When $r_2>0$, $r_1<0$ (but close to $T_{c,1}$), we can approximate $\varphi \sim -\frac{\sqrt{3} v e M }{r_2}$, where $e=e_{xy}=e_{zx}=e_{yz}$ quantifies the strength of external strain. After substituting this back in the free energy we find the following shift in the transition temperature of SI order
\begin{equation}
T_{c,1}^{\prime}=T_{c,1}+\frac{3v^2 e^2}{r_{01}r_{02}(T_{c,1}-T_{c,2})},\label{strain3}
\end{equation}
when $T_{c,2} << T_{c,1}$. However, the energy scale of external strain should be smaller than the transition temperature of the dominant order say $T_{c,1}$. 

This prediction should be relevant for experiments on thin films of Pr$_2$Ir$_2$O$_7$, as growth process of thin films along [1,1,1] direction naturally introduces $T_{2g}$ strain. From Eq.~(\ref{tempHall}) we naturally find the following relation
\begin{equation}
\frac{T_H(w_1)}{T_H(w_2)}=\left(\frac{\sigma_H(w_1)}{\sigma_H(w_2)}\right)^2,\label{thinfilmHall}
\end{equation}
between the thickness ($w$) dependent anomalous Hall conductivities and the corresponding onset temperatures [$T_H(w)$]. Therefore the onset temperature can increase by an order of magnitude even when the anomalous Hall conductivity only changes by a factor $\mathcal{O}(1)$ (say $3$ to $4$), which is an important prediction of our theory. Next we consider the role of externally applied magnetic field.

\subsection{External magnetic field}\label{magfield}

In addition to breaking time reversal symmetry, an external magnetic field ($\mathbf{H}$) also reduces the cubic symmetry. Therefore, it can cause coupling among different types of multipolar order parameters. Since the SI order parameter $M_a$ is built out of both dipolar ($J_a$) and octupolar ($J^3_a$) quantities, it can directly couple to both $H_a$ and $H^3_a$. By contrast, a purely octopular AIAO order can only couple to the product $H_1H_2H_3$. In addition, products like $H_aH_b$ can couple to rotational symmetry breaking but time reversal symmetric quadrupolar order parameters $Q_{ab}$. Finally the third order terms $H_1(H^2_2-H^2_3)$, $H_2(H^2_3-H^2_1)$ and $H_3(H^2_1-H^2_2)$ can respectively couple to the time reversal symmetry breaking $T_{2u}$ octupolar order parameters $T_{2u,1}=\langle \Psi^\dagger J_1(J^2_2-J^2_3) \Psi \rangle$, $T_{2u,2}=\langle \Psi^\dagger J_2(J^2_3-J^2_1) \Psi \rangle$ and $T_{2u,3}=\langle \Psi^\dagger J_3(J^2_1-J^2_2) \Psi \rangle$. \emph{By coupling to higher order terms in magnetic field strength, the competing multipolar orders will generally affect the nonlinear susceptibility measurements in torque magnetometry}~\cite{LiangOng}.

For $\mathbf{H}=H[1,1,1]$, the free energy density can support the following additional terms
\begin{eqnarray}
f_{[111]}&=&\sum_{a=1}^{3}[\lambda_1 \; H M_a + \lambda_2 \; H^3 M^3_a]+ \lambda_3 \; H^3 \varphi \nonumber \\ &+&\lambda_4 H^2\; \sum_{a<b=1}^{3} Q_{ab}+ \lambda_5  H^2 \varphi \; \sum_{a=1}^{3} M_a,
\end{eqnarray}
where $\lambda_j$s are appropriate coupling constants. Consequently, a magnetic field along [1,1,1] direction is capable of inducing  a general form of 3I1O order and $T_{2g}$ nematicity. Such effects will be significant when the measurements are performed on Nd$_{2-2x}$Pr$_{2x}$Ir$_2$O$_7$, but either on the metallic side $x>0.8$ or in the close proximity of the MIT on the insulating side. Important effects of competing orders should also be observable in field dependent measurements on Nd$_2$Ir$_2$O$_7$ under hydrostatic pressure when $p \sim p_c$. When we are deep inside the AIAO ordered phase, such effects will be very small and only a strong magnetic field will induce a MIT.

By contrast, a field along [0,0,1] direction (or [1,0,0], [0,1,0]) couples only to the uniaxial SI order parameter $M_3$ (or $M_1$, $M_2$) and the $E_{g}$ quadrupolar order parameter $Q=(Q_{11}+Q_{22}-Q_{33})/4=\langle \Psi^\dagger J^2_3 \Psi \rangle \propto \langle \Psi^\dagger \Gamma_5 \Psi \rangle$ according to
\begin{eqnarray}
f_{[001]}&=&[\lambda_1 H M_3 + \lambda_2H^3 M^3_3]+g_4 H^2 Q.
\end{eqnarray}
Therefore, a sufficiently strong field along the [0,0,1] direction can overcome the cubic anisotropy, and lead to the nucleation of uniaxial SI order~\cite{Ueda2,Nakatsuji7}, together with $E_{g}$ nematicity on the metallic side of the global phase diagram. Since AIAO order does not couple to [0,0,1] field direction, it is easier to suppress the AIAO order for such a field orientation. This latter effect has been clearly demonstrated through a field dependent MIT observed in Nd$_2$Ir$_2$O$_7$~\cite{Ueda2,Nakatsuji7}.

Finally an external field $\mathbf{H}=H[1,1,0]$ modifies the free energy through the terms
\begin{eqnarray}
f_{[110]}&=&\sum_{a=1}^{2}[\lambda_1 H M_a +\lambda_2 H^3 M^3_a]+ \lambda_5 H^2 M_3 \varphi \nonumber \\ &+& \lambda_4 H^2 \; Q_{12}+\lambda_6 H^3 \left(T_{2u,1}- T_{2u,2}\right), \nonumber \\
\end{eqnarray}
Based on such couplings, we expect a complicated form of induced multipolar orders in the presence of sufficiently strong magnetic field along [1,1,0] direction on the metallic side of the global phase diagram. However due to the presence of several competing orders it can be difficult to induce MIT when we are deep inside insulating phase. It is worth pointing out that no MIT has been induced in Nd$_2$Ir$_2$O$_7$~\cite{Ueda2,Nakatsuji7} for field along [1,1,0] direction. Therefore experiments for this field orientation has to be carried out either on the metallic side or in the close proximity of the MIT for observing any significant effect. Any further analysis of the effects of external magnetic field is left for future investigation.

\subsection{Additional experimental implications}

The generic phase diagram of pyrochlore iridates, shown in Figs.~\ref{fig:2d} and \ref{fig:2e}, can be experimentally verified through the measurement of anomalous Hall conductivity ($\sigma_H$) on Nd$_{2-2x}$Pr$_{2x}$Ir$_2$O$_7$. As the concentration of Nd is gradually increased in this compound, we expect the strength of AHE to gradually decrease. Finally across a critical Nd-doping ($x_c=0.8$) the system is expected to enter into the AIAO phase that does not support any AHE~\cite{Ueda3}. Whether $\sigma_H$ vanishes continuously (Fig.~\ref{fig:2d}) or discontinuously (Fig.~\ref{fig:2e}) can distinguish these two possibilities. 

On the other hand the metallic phase for metal-insulator QPT induced by hydrostatic pressure can in principle correspond to either (i) a PSM or (ii) 3I1O state. These two situations can be distinguished based on the absence or presence of AHE. Therefore measurement of AHE or magnetooptical properties are required at $p>p_c$ to determine the actual nature of the metallic state obtained through hydrostatic pressure.

The precise nature of Weyl fermion can be probed by carrying out Fourier transformed STM measurements (since $T_H \sim 1.5 K$, one cannot perform ARPES). This will allow detection of surface states, also known as Fermi arcs. However, our proposal for enhancing $T_H$ in thin films by an order of magnitude will enable direct probing of quasiparticles through ARPES measurements. We also note that the difference in nodal topology also manifests in the density of states for the low energy quasiparticles~\cite{Bera}, which generally lead to distinct temperature and frequency dependent response functions for the symmetry breaking metallic state. 

Perhaps the best probe of magnetic ordering is neutron scattering. However, Ir is a very good absorber of neutrons. Therefore, neutron scattering is not an efficient probe for magnetic ordering of Ir electrons. In some of the AIAO ordered materials, the measured signal comes from lanthanide ions at a temperature considerably lower than the actual ordering temperature. However, muon resonance can provide valuable insight regarding the magnetic nature of enigmatic time reversal symmetry breaking phase in Pr$_2$Ir$_2$O$_7$. It is worth mentioning that the muon resonance technique has been employed for probing the AIAO magnetic order in Eu$_2$Ir$_2$O$_7$~\cite{muon-1}. 
   
\section{Conclusion}\label{conclusion}

In contrast to most prior theoretical works on Pr$_2$Ir$_2$O$_7$ mainly focusing on weak exchange (super exchange or RKKY) coupling among Pr local moments, we have eschewed the local moment magnetism in favor of itinerant magnetic ordering. We have modeled the normal state of 227 iridates as a parabolic semimetal in the presence of generic short range interactions (see Sec.~\ref{competingorder}), and classified the possible momentum independent multipolar orders according to the representation of cubic symmetry. The generic form of local interactions has been described in terms of six effective coupling constants. For constructing the global phase diagram of an interacting parabolic semimetal, we have used the following three complementary theoretical methods (i) perturbative one loop renormalization group analysis, (ii) mean-field theory of ordered state, and (iii) Landau theory of competing order parameters. Based on one loop RG analysis of interaction couplings and order parameter susceptibilities, we identify the dominant ordering tendencies of the semimetallic normal state in an unbiased manner, as the renormalization group flow accounts for the fluctuation effects of all incipient ordering channels on an equal footing (see SubSection~\ref{subsecRG}). The renormalization group calculations reveal that the parabolic semimetal is stable against weak interactions, and the onset of any ordering occurs through a continuous quantum phase transition. The itinerant quantum critical points are non-Gaussian in nature. After isolating the dominant ordering channel, we apply the mean-field theory for describing the reconstructed band structure and associated nodal topology (see Sec.~\ref{AHE}). Within the framework of a phenomenological Landau theory, we finally address the transition between two competing ordered states, which provides valuable insight into the structure of phase diagram in the strong interaction regime (Sec.~\ref{landautheory}). The generic effective model and its RG analysis as described in this work can also be applied to other correlated materials with quadratic band touching. We also note that in the presence of an underlying Fermi surface, one may have a fluctuation driven first order quantum phase transition between spin-ice and paramagnetic phases~\cite{belitz-kirkpatrick}.

Motivated by the phenomenology of 227 iridates we have mainly considered the competition between ``all-in all-out" (pure octupolar) and ``spin ice" (mixture of dipolar and octupolar states) ordered states. On a general symmetry ground [Sec.~\ref{landautheory}], we have showed that the cubic environment locks the vector order parameter $\mathbf{M}$ for ``spin ice" phase along one of the [1,1,1] directions, giving rise to a ``triplet spin ice" phase, where the Ir electrons show ``3-in-1-out" configurations for the expectation value of the spin operators. We have also showed that due to the combined effects of cubic anisotropy and particle hole asymmetry of the normal state, ``3-in-1-out" phase generally becomes an admixture of SI and AIAO order parameters. 

We have discussed why any time reversal symmetry breaking order is capable of producing Weyl quasiparticles (see Sec.~\ref{AHE}, and  Fig.~\ref{nodalband}, Fig.~\ref{berry1} and Fig.~\ref{berry2}). Due to the absence of cubic symmetry, Weyl excitations of ``spin ice" order causes anomalous Hall conductivity $\sigma_H \sim \sqrt{|\mathbf{m}|}$, where $\mathbf{m}$ is the magnetic moment per Ir ion [see Eqs.~(\ref{HallSI}), (\ref{AH111}), (\ref{AHE111})]. We have showed that the ``3-in-1-out" order gives rise to a Weyl metal with two Weyl points and a large anomalous Hall effect along one of the eight possible [1,1,1] directions without any large magnetic moment. The experimentally observed direction, size, and onset temperature of anomalous Hall conductivity can be reconciled with an amplitude of ``3-in-1-out" order, which gives rise to an immeasurably small magnetic moment $\sim 10^{-5} \mu_B$ per Ir ion [see Sec.~\ref{section-pr}]. The Weyl metal phase can also support large anomalous thermal Hall and Nernst effects.

We have explicitly demonstrated why ``3-in-1-out" order generally causes weak electronic nematicity leading to a small lattice distortion along [1,1,1] direction [see Eqs.~(\ref{strain1}), (\ref{strain2})]. Based on this we have showed that an externally applied weak strain along [1,1,1] direction leads to an enhancement of AHE, making it observable at relatively higher temperatures [see Eqs.~(\ref{strain3}) and (\ref{thinfilmHall})]. Therefore, an order of magnitude enhancement of $T_H$ would result in a minor increase in $\sigma_H$ by a factor of unity (say 3 or 4), since $T_H \propto \sigma^2_H$. Hence future anomalous Hall effect and magnetooptical measurements on thin films of Pr$_2$Ir$_2$O$_7$ (which naturally inherit strain along [1,1,1] direction during the growth process) can serve as an important test of our theory.

Whether the transition between ``3-in-1-out" and ``all-in all-out" ordered states is second or first order in nature can be pinned down by systematically measuring $\sigma_H$ in Nd$_{2-2x}$Pr$_{2x}$Ir$_2$O$_7$, where $x$ controls the chemical pressure and there is a metal-insulator quantum phase transition around $x \sim 0.8$. We have also discussed an intriguing possibility of non-Gaussian itinerant quantum phase transition between the parabolic semimetal and ``3-in-1-out" ordered phase, which can be accessed either by applying hydrostatic pressure on Pr$_2$Ir$_2$O$_7$ or chemical pressure through substitution of Pr by larger Ce or La ions. It would also be interesting to replace nonmagnetic Eu ion of ``all-in all-out" ordered insulating compound Eu$_2$Ir$_2$O$_7$ with larger, nonmagnetic La ion to induce a metal-insulator transition. Whether the emergent metallic phase breaks time reversal symmetry (displaying anomalous Hall effect) or not would provide valuable experimental insight into the global phase diagram. Due to  the strong fluctuations among eight-fold degenerate ``3-in-1-out" states, the actual ground state can remain disordered even below the mean-field transition temperature $T_H$ due to the lack of spin-stiffness. Thus, we have developed a fairly complete phenomenological Landau theory describing the generic phase diagrams of 227 pyrochlore iridates, arguing that many of the observed anomalous properties of Pr$_2$Ir$_2$O$_7$ can be understood from an underlying itinerant ``3-in-1-out" magnetic order and emergent Weyl excitations without invoking any large local magnetic moment and associated Kondo physics. We would like to emphasize that we are finding the itinerant versions of ``all-in all-out", ``2-in-2-out" and ``3-in-1-out" magnetic configurations for Ir ions, as a consequence of sufficiently strong and generic local interactions among delocalized Ir electrons. For insulating pyrochlore systems, such orderings are often found by considering strongly anisotropic magnetic interactions among local moments. The additional effects of magnetic anisotropy for some of the lanthanide ions (since not all lanthanide ions are magnetic) on our proposed phase diagrams will be an interesting topic of future research.   

\acknowledgements This work is supported by JQI-NSF-PFC and LPS-MPO-CMTC. We thank N. P. Armitage, L. Balicas, A. A. Burkov, I. F. Herbut, Y. B. Kim, R. Moessner, S. Nakatsuji, A. M. Turner, and L. Wu for helpful discussions. B. R. thanks Aspen Center for Physics for hospitality during the Summer Program (2015).

\onecolumngrid

\appendix

\section{Derivation of Luttinger Hamiltonian} \label{sec:appendix-A}

We first outline the methodology for projecting out the higher energy bands for obtaining low energy Luttinger Hamiltonian for noninteracting electrons. In the momentum space the tight-binding Hamiltonian of Eq.~(\ref{eqtb}) can be compactly written by using $8 \times 8$ $SU(8)$ generators, obtained by taking direct products of $2 \times 2$ Pauli matrices $\sigma_{\mu}$ (operating on spin indices) and $4\times4$ $SU(4)$ generators $\hat{\lambda}_j$s (operating on orbital indices). After projecting out two fold degenerate conduction bands, the model can be written in term of $6 \times 6$ $SU(6)$ generators, found by taking direct products of $2 \times 2$ Pauli matrices $\sigma_{\mu}$ and $3 \times 3$ $SU(3)$ generators $\hat{t}_j$s. After accounting for spin-orbit splitting we will arrive at the subspace of quadratically touching bands, where the effective Hamiltonian can be expressed in terms of $4 \times 4$ $SU(4)$ generators or Dirac matrices.

The spin independent and spin dependent parts of the tight-binding model respectively become
\allowdisplaybreaks[4]
\begin{eqnarray}
H_0&=&\sum_{\mathbf{k}}\chi^\dagger_{\mathbf{k}}\hat{H}_0 (\mathbf k)\chi_{\mathbf{k}}= \sum_{\mathbf{k}}\chi^\dagger_{\mathbf{k}} \; \sigma_0 \otimes \bigg[\hat{\lambda}_1 F_1(\mathbf{k})+\hat{\lambda}_4 F_2(\mathbf{k})+\hat{\lambda}_6 F_3(\mathbf{k})+\hat{\lambda}_9 F_4(\mathbf{k})+\hat{\lambda}_{11} F_5(\mathbf{k}) +\hat{\lambda}_{13} F_6(\mathbf{k})\bigg] \chi_{\mathbf{k}}, \\
H_{SO}&=&\sum_{\mathbf{k}}\chi^\dagger_{\mathbf{k}}\hat{H}_{SO} (\mathbf k) \chi_{\mathbf{k}}= i \; \sqrt{2} \xi \; \sum_{\mathbf{k}} \chi^\dagger_{\mathbf{k}} \bigg[-(\sigma_1+\sigma_2) \otimes \hat{\lambda}_2 C_{12}^{-} +(\sigma_2+\sigma_3) \otimes \hat{\lambda}_5 C_{23}^{-} -(\sigma_3+\sigma_1) \otimes \hat{\lambda}_7 C_{31}^{-} \nonumber \\
&&-(\sigma_1 -\sigma_2) \otimes \hat{\lambda}_{14} C_{12}^{+}-(\sigma_2 -\sigma_3) \otimes \hat{\lambda}_{12} C_{23}^{+}-(\sigma_3 -\sigma_1) \otimes \hat{\lambda}_{10} C_{31}^{+}\bigg]\chi_{\mathbf{k}},
\end{eqnarray}
where $\chi_{\mathbf{k}}$ is an eight component spinor comprising of the fermion annihilation operators $c_{i,s,\mathbf{k}}$, with $i=1,2,3,4$ representing the four sites of a tetrahedron, and spin projection $s=\uparrow /\downarrow$. The momentum dependent functions are defined as $C_{ij}^{\pm}=\cos (k_i \pm k_j)$ and
\begin{eqnarray}
F_1(\mathbf{k})=2t_1 C_{12}^{-}+4t_2 \; \cos(2k_3)C_{12}^{+}, \quad
F_2(\mathbf{k})=2t_1 C_{23}^{-}+4t_2 \; \cos(2k_1)C_{23}^{+}, \quad
F_3(\mathbf{k})=2t_1 C_{31}^{-}+4t_2 \; \cos(2k_2)C_{31}^{+}, \nonumber \\
F_4(\mathbf{k})=2t_1 C_{12}^{+}+4t_2 \; \cos(2k_3)C_{12}^{-}, \quad
F_5(\mathbf{k})=2t_1 C_{23}^{+}+4t_2 \; \cos(2k_1)C_{23}^{-}, \quad
F_6(\mathbf{k})=2t_1 C_{31}^{+}+4t_2 \; \cos(2k_2)C_{12}^{-}. \nonumber \end{eqnarray}
The $SU(4)$ generators are
\begin{eqnarray}
&&\hat{\lambda}_1=  \begin{bmatrix}
    0 & 1 & 0 & 0  \\
    1 & 0 & 0 & 0  \\
    0 & 0 & 0 & 0 \\
    0 & 0 & 0 & 0  \\
  \end{bmatrix}, \; \hat{\lambda}_2=  \begin{bmatrix}
    0 & -i & 0 & 0  \\
    i & 0 & 0 & 0  \\
    0 & 0 & 0 & 0  \\
    0 & 0 & 0 & 0  \\
  \end{bmatrix}, \; \hat{\lambda}_3=  \begin{bmatrix}
    1 & 0 & 0 & 0  \\
    0 & -1 & 0 & 0  \\
    0 & 0 & 0 & 0  \\
    0 & 0 & 0 & 0  \\
  \end{bmatrix},
	\hat{\lambda}_4=  \begin{bmatrix}
    0 & 0 & 1 & 0 \\
    0 & 0 & 0 & 0 \\
    1 & 0 & 0 & 0  \\
    0 & 0 & 0 & 0 \\
  \end{bmatrix}, \; \hat{\lambda}_5=  \begin{bmatrix}
    0 & 0 & -i & 0 \\
    0 & 0 & 0 & 0 \\
    i & 0 & 0 & 0  \\
    0 & 0 & 0 & 0 \\ \end{bmatrix}, \nonumber \\ 		
		&&\hat{\lambda}_6=  \begin{bmatrix}
    0 & 0 & 0 & 0 \\
    0 & 0 & 1 & 0 \\
    0 & 1 & 0 & 0  \\
    0 & 0 & 0 & 0 \\
  \end{bmatrix}, \; \hat{\lambda}_7=  \begin{bmatrix}
    0 & 0 & 0 & 0 \\
    0 & 0 & -i & 0 \\
    0 & i & 0 & 0  \\
    0 & 0 & 0 & 0 \\
  \end{bmatrix} , \; \hat{\lambda}_8= \frac{1}{\sqrt{3}} \begin{bmatrix}
    1 & 0 & 0 & 0 \\
    0 & 1 & 0 & 0 \\
    0 & 0 & -2 & 0  \\
    0 & 0 & 0 & 0 \\
  \end{bmatrix} , \; \hat{\lambda}_9= \begin{bmatrix}
    0 & 0 & 0 & 1 \\
    0 & 0 & 0 & 0 \\
    0 & 0 & 0 & 0  \\
    1 & 0 & 0 & 0 \\
  \end{bmatrix} , \; \hat{\lambda}_{10}= \begin{bmatrix}
    0 & 0 & 0 & -i \\
    0 & 0 & 0 & 0 \\
    0 & 0 & 0 & 0  \\
    i & 0 & 0 & 0 \\
  \end{bmatrix} , \nonumber \\  && \hat{\lambda}_{11}= \begin{bmatrix}
    0 & 0 & 0 & 0 \\
    0 & 0 & 0 & 1 \\
    0 & 0 & 0 & 0  \\
    0 & 1 & 0 & 0 \\
  \end{bmatrix}, \; \hat{\lambda}_{12}= \begin{bmatrix}
    0 & 0 & 0 & 0 \\
    0 & 0 & 0 & -i \\
    0 & 0 & 0 & 0  \\
    0 & i & 0 & 0 \\
  \end{bmatrix}, \; \hat{\lambda}_{13}= \begin{bmatrix}
    0 & 0 & 0 & 0 \\
    0 & 0 & 0 & 0 \\
    0 & 0 & 0 & 1  \\
    0 & 0 & 1 & 0 \\
  \end{bmatrix} , \; \hat{\lambda}_{14}= \begin{bmatrix}
    0 & 0 & 0 & 0 \\
    0 & 0 & 0 & 0 \\
    0 & 0 & 0 & -i  \\
    0 & 0 & i & 0 \\
  \end{bmatrix} , \; \hat{\lambda}_{15}= \frac{1}{\sqrt{6}}\begin{bmatrix}
    1 & 0 & 0 & 0 \\
    0 & 1 & 0 & 0 \\
    0 & 0 & 1 & 0  \\
    0 & 0 & 0 & -3 \\
  \end{bmatrix}. \nonumber \\
\end{eqnarray}
In the absence of spin orbit coupling ($\xi=0$), there is a six-fold degeneracy at the $\Gamma$ point. This can be seen by diagonalizing $\hat{H}_0 (\mathbf{ k}=0)$ through the unitary transformation
$S^{\dagger} \hat{H}_0(k=0) S =-\left(2t_1+4t_2\right) \sqrt{6} \sigma_0 \otimes \hat{\lambda}_{15}$ with
\begin{eqnarray}
S^{\dagger}=S= \frac{1}{2}\sigma_0 \otimes \begin{bmatrix}
    1 & -1 & 1 & -1 \\
    -1 & 1 & 1 & -1 \\
    1 & 1 & -1 & -1  \\
    -1 & -1 & -1 & -1 \\
  \end{bmatrix}=\frac{1}{2}\sigma_0 \otimes \left[-\hat{\lambda}_1+\hat{\lambda}_4+\hat{\lambda}_6-\hat{\lambda}_9-\hat{\lambda}_{11}-\hat{\lambda}_{13}+2\sqrt{3} \; \hat{\lambda}_8+\sqrt{6} \; \hat{\lambda}_{15}\right].
\end{eqnarray}
By expanding $\hat{H}_0$ up to the quadratic order, we obtain
\begin{eqnarray}
S^\dagger \hat{H}_0 (\mathbf k) S = -\sqrt{6}\left[2b-(a+2b)\frac{k^2}{3}\right]\sigma_0 \otimes \hat{\lambda}_{15}&-&\frac{c}{\sqrt{3}}\sigma_0 \otimes \left[d_1(\hat{\lambda}_6-\hat{\lambda}_9)+d_2(\hat{\lambda}_4-\hat{\lambda}_{11})+d_3(\hat{\lambda}_1-\hat{\lambda}_{13})\right]\nonumber \\
&-&\frac{2}{\sqrt{3}}(b-a)\; \sigma_0 \otimes \; \left[d_4 \hat{\lambda}_3-d_5\hat{\lambda}_{8}\right] + \mathcal{O}(k^4_j),
\end{eqnarray}
where $a=8 t_2$, $b=t_1+2 t_2$ and $c=2(t_1-2 t_2)$, and we have used five normalized $d$-wave form factors $d_1=-\sqrt{3}k_yk_z$, $d_2=-\sqrt{3}k_zk_x$, $d_3=-\sqrt{3}k_xk_y$, $d_4=-\frac{\sqrt{3}}{2}(k^2_x-k^2_y)$ and $d_5=-\frac{1}{2}(2k^2_z-k^2_x-k^2_y)$. At this stage we project out the non-degenerate band and work with the six-fold degenerate subspace, and the projected form of $S^\dagger \hat{H}_0 (\mathbf k) S$ becomes
\begin{eqnarray}
\hat{H}^{\prime}(\mathbf k)=-\left( 2 b - (a+2 b) \frac{k^2}{3}\right) \sigma_0 \otimes \mathbb{1}_{3 \times 3}  -
\bigg[ \frac{c}{\sqrt{3}} \; \sigma_0 \otimes \left(d_1  \hat{t}_6 + d_2 \hat{t}_4+d_3 \hat{t}_1 \right) + \frac{2(b-a)}{\sqrt{3}} \; \sigma_0 \otimes\left(d_4 \hat{t}_3-d_5 \hat{t}_8 \right)\bigg],
\end{eqnarray}
and the $3 \times 3$ $SU(3)$ Gell-Mann matrices are given by
\allowdisplaybreaks[4]
\begin{eqnarray}
&&\hat{t}_1=  \begin{bmatrix}
    0 & 1 & 0  \\
    1 & 0 & 0  \\
    0 & 0 & 0 \\
    \end{bmatrix}, \: \hat{t}_2=  \begin{bmatrix}
    0 & -i & 0 \\
    i & 0 & 0 \\
    0 & 0 & 0  \\
    \end{bmatrix}, \: \hat{t}_3=  \begin{bmatrix}
    1 & 0 & 0  \\
    0 & -1 & 0  \\
    0 & 0 & 0  \\
    \end{bmatrix}, \: \hat{t}_4=  \begin{bmatrix}
    0 & 0 & 1  \\
    0 & 0 & 0 \\
    1 & 0 & 0  \\
     \end{bmatrix}, \: \nonumber \\
		&& \hat{t}_5=  \begin{bmatrix}
    0 & 0 & -i \\
    0 & 0 & 0 \\
    i & 0 & 0  \\
    \end{bmatrix}, \: \hat{t}_6=  \begin{bmatrix}
    0 & 0 & 0 \\
    0 & 0 & 1 \\
    0 & 1 & 0 \\
    \end{bmatrix}, \: \hat{t}_7=  \begin{bmatrix}
    0 & 0 & 0 \\
    0 & 0 & -i \\
    0 & i & 0 \\
    \end{bmatrix} , \: \hat{t}_8= \frac{1}{\sqrt{3}} \begin{bmatrix}
    1 & 0 & 0 \\
    0 & 1 & 0 \\
    0 & 0 & -2 \\
    \end{bmatrix}.
    \end{eqnarray}

The spin-orbit coupling reduces the six-fold degeneracy and produces a two fold degenerate band, in addition to quadratically touching Kramers degenerate conduction and valence bands. For obtaining the appropriate low energy description we diagonalize the projected spin-orbit term at the $\Gamma$ point with the unitary matrix
\begin{eqnarray}
U=\begin{bmatrix}
-\frac{1}{\sqrt{2}} \sigma_3 & \frac{i}{\sqrt{6}} \sigma_2 & -\frac{1}{\sqrt{3}} \sigma_1 \\
-\frac{i}{\sqrt{2}} \sigma_0 & - \frac{i}{\sqrt{6}} \sigma_1 & -\frac{1}{\sqrt{3}} \sigma_2 \\
\hat{O} &  \frac{2}{\sqrt{6}} \sigma_0 & -\frac{1}{\sqrt{3}} \sigma_3 \\
\end{bmatrix},
\end{eqnarray}
where $\hat{O}$ is a two dimensional null matrix, and find $\hat{U}^\dagger \hat{H}_{SO}(\mathbf{k}=0) U= 2\sqrt{6} \; \xi \; \hat{t}_8$. By projecting on to the subspace of quadratically touching bands we arrive at the following $4 \times 4$ low energy Hamiltonian
\begin{eqnarray}
H=\int \frac{d^3k}{(2\pi)^3} \psi^{\dagger}_{\mathbf{k}}\bigg[- \left( 2 b -2\sqrt{2} \xi- (a+2 b) \frac{k^2}{3}\right) \mathbb{1}_{4\times4}- \frac{c}{3} \left(d_1 \tilde{\Gamma}_1+d_2 \tilde{\Gamma}_2+d_3 \tilde{\Gamma}_3 \right)-\frac{2}{3} (b-a) \left(d_4 \tilde{\Gamma}_4+d_5 \tilde{\Gamma}_5 \right)\bigg]\psi^\dagger_{\mathbf{k}},
\end{eqnarray}
where
\begin{eqnarray}
\tilde{\Gamma}_1= \begin{bmatrix}
\hat{O} & -i \sigma_0 \\
i \sigma_0 & \hat{O} \\
\end{bmatrix}, \:
\tilde{\Gamma}_2= \begin{bmatrix}
\hat{O} & \sigma_3 \\
\sigma_3 & \hat{O} \\
\end{bmatrix}, \:
\tilde{\Gamma}_3= \begin{bmatrix}
\hat{O} & \sigma_2 \\
\sigma_2 & \hat{O} \\
\end{bmatrix}, \:
\tilde{\Gamma}_4= \begin{bmatrix}
\hat{O} & \sigma_1 \\
\sigma_1 & \hat{O} \\
\end{bmatrix}, \:
\tilde{\Gamma}_5= \begin{bmatrix}
\sigma_0 & \hat{O}  \\
 \hat{O} & -\sigma_0 \\
\end{bmatrix},
\end{eqnarray}
are five mutually anti-commuting matrices. After performing yet another unitary transformation $\tilde{U}^\dagger \psi= \Psi$ with
\begin{equation}
\tilde{U}= \begin{bmatrix}
1 & 0 & 0 & 0 \\
0 & 0 & 0 & 1 \\
0 & 1 & 0 & 0 \\
0 & 0 & 1 & 0 \\
\end{bmatrix}
\end{equation}
we obtain the canonical form of the Luttinger Hamiltonian (after restoring the factor of $\hbar$)
\begin{equation}
H=\int \frac{d^3k}{(2\pi)^3} \Psi^\dagger_{\mathbf{k}}\hat{H}_L(\mathbf k) \Psi_{\mathbf{k}}=\int \frac{d^3k}{(2\pi)^3} \Psi^\dagger_{\mathbf{k}}\bigg[ \left( E_0 + \frac{\hbar^2 k^2}{2 m_0}\right) \mathbb{1}_{4 \times 4}- \frac{\hbar^2}{2m_1} \left(d_1 \Gamma_1+d_2 \Gamma_2+d_3 \Gamma_3 \right)-\frac{\hbar^2}{2m_2}  \left(d_4 \Gamma_4+d_5 \Gamma_5 \right)\bigg]\Psi_{\mathbf{k}}.
\end{equation}
The new set of five mutually anti-commuting $\Gamma$ matrices are defined as
\begin{eqnarray}
\Gamma_1&=&\frac{J_2 J_3+J_3 J_2}{\sqrt{3}}= \begin{bmatrix}
\sigma_2 & \hat{O} \\
\hat{O} & -\sigma_2 \\
\end{bmatrix}, \:
\Gamma_2= \frac{J_3 J_1+J_1 J_3}{\sqrt{3}}=\begin{bmatrix}
\sigma_3 & \hat{O} \\
\hat{O} & -\sigma_3 \\
\end{bmatrix}, \:
\Gamma_3=\frac{J_1 J_2+J_2 J_1}{\sqrt{3}}= \begin{bmatrix}
\hat{O} & -i \sigma_0 \\
i \sigma_0 & \hat{O} \\
\end{bmatrix}, \nonumber \\
\Gamma_4&=&\frac{1}{\sqrt{3}} \left(J^2_1-J^2_2 \right)= \begin{bmatrix}
\hat{O} &  \sigma_0 \\
\sigma_0 & \hat{O} \\
\end{bmatrix}, \:
\Gamma_5=\frac{1}{3} \left(2J^2_3-J^2_1-J^2_2 \right)= \begin{bmatrix}
\sigma_3 & \hat{O}  \\
 \hat{O} & -\sigma_3 \\
\end{bmatrix},
\end{eqnarray}
where $J_a$ are spin-3/2 matrices.

The spin density $\Psi^\dagger \mathbf{J} \Psi$ transforms as a dipolar object following $T_{1u}$ representation, and in terms of the $\Gamma$ matrices we can write
\begin{eqnarray}
&& \Psi^\dagger J_1 \Psi=\Psi^\dagger\left[\frac{\sqrt{3}}{2} \Gamma_{15}-\frac{1}{2} \left( \Gamma_{23}-\Gamma_{14}\right)\right]\Psi, \quad
\Psi^\dagger J_2 \Psi=\Psi^\dagger\left[-\frac{\sqrt{3}}{2} \Gamma_{25}-\frac{1}{2} \left( \Gamma_{13}+\Gamma_{24}\right)\right]
\Psi, \nonumber \\
&& \Psi^\dagger J_3 \Psi=\Psi^\dagger\left[-\Gamma_{34}- \frac{1}{2} \Gamma_{12}\right]\Psi,\label{dipole}
\end{eqnarray}
where $\Gamma_{jk}=[\Gamma_j,\Gamma_k]/(2 i)$. By contrast, $\Psi^\dagger J^3_a \Psi$ transform as octupolar quantities following $T_{1u}$ representation and
\begin{eqnarray}
&& \Psi^\dagger J^3_1 \Psi=\Psi^\dagger\left[\frac{7\sqrt{3}}{8} \Gamma_{15} +\frac{7}{8}\Gamma_{14}-\frac{13}{8} \Gamma_{23}\right]\Psi, \quad
\Psi^\dagger J^3_2 \Psi=\Psi^\dagger\left[-\frac{7\sqrt{3}}{8} \Gamma_{25} +\frac{7}{8}\Gamma_{24}+\frac{13}{8} \Gamma_{13}\right]\Psi, \nonumber \\
&& \Psi^\dagger J^3_3 \Psi=\Psi^\dagger\left[-\frac{13}{8}\Gamma_{12}-\frac{7}{4} \Gamma_{34}\right]\Psi.\label{octupoleT1u}
\end{eqnarray}
Finally the octupolar quantities following $A_{2u}$ and $T_{2u}$ representation are respectively given by
\begin{eqnarray}
&& \Psi^\dagger \left[J_1 J_2 J_3+J_3 J_2 J_1\right]\Psi=-\frac{\sqrt{3}}{2} \Psi^\dagger \Gamma_{45} \Psi, \label{octupoleA2u}\\
&& \Psi^\dagger \left\{ J_1,J^2_2-J^2_3 \right\}\Psi=\Psi^\dagger\left[-\frac{\sqrt{3}}{2} \Gamma_{15}+\frac{3}{2} \Gamma_{14}\right]\Psi, \quad
\Psi^\dagger \left\{ J_2,J^2_3-J^2_1 \right\} \Psi=\Psi^\dagger\left[-\frac{\sqrt{3}}{2} \Gamma_{25}-\frac{3}{2} \Gamma_{24}\right]\Psi, \nonumber \\
&& \Psi^\dagger \left\{ J_3,J^2_1-J^2_2 \right\} \Psi=\sqrt{3}\Psi^\dagger \Gamma_{35} \Psi.\label{octupoleT2u}
\end{eqnarray}

\section{Coupling between magnetic order parameter and itinerant fermion} \label{sec:appendix-B}

\emph{All-in all-out}: The AIAO order couples to underlying fermions according to $\varphi \; \chi^\dagger \hat{O}_{0} \chi$ where the $8 \times 8$ matrix
\begin{eqnarray}
\hat{O}_0=\frac{1}{\sqrt{3}} \; \mathrm{diag} \bigg[ \sigma_1-\sigma_2+\sigma_3, -\sigma_1+\sigma_2+\sigma_3, \sigma_1+\sigma_2-\sigma_3, -\sigma_1-\sigma_2-\sigma_3 \bigg].
\end{eqnarray}
After projecting onto the low energy subspace of three bands, $\hat{O}_0$ modifies into the following $6 \times 6$ matrix
\begin{eqnarray}
\hat{O}^\prime_0=\frac{1}{\sqrt{3}} \; \bigg[ \sigma_1 \otimes \hat{t}_6+ \sigma_2 \otimes \hat{t}_4+\sigma_3 \otimes \hat{t}_1 \bigg].
\end{eqnarray}
After the subsequent unitary transformation we find
\begin{eqnarray}
U^\dagger \hat{O}^\prime_0 U= \sigma_1 \otimes \hat{t}_2=\begin{bmatrix}
\hat{O} & - i \sigma_1 & \hat{O} \\
i \sigma_1 & \hat{O} & \hat{O} \\
\hat{O} & \hat{O} & \hat{O} \\
\end{bmatrix},
\end{eqnarray}
where $\hat{O}$ is $2\times 2$ null matrix. Finally performing the unitary rotation with $\tilde{U}$ and comparing with Eq.~(\ref{octupoleA2u}) we see that the AIAO is an example of $A_{2u}$ octupolar order and it couples to the quadratically touching bands according to $\varphi \Psi^\dagger \Gamma_{45} \Psi$.

\emph{Spin ice}: Next we consider a particular 2I2O or SI configuration that couples with underlying fermions according to $M_3 \;\chi^\dagger \hat{O}_{3} \chi$ where the $8 \times 8$ matrix
\begin{eqnarray}
\hat{O}_3=\frac{1}{\sqrt{3}} \; \mathrm{diag} \bigg[ -\sigma_1+\sigma_2-\sigma_3, \sigma_1-\sigma_2-\sigma_3, \sigma_1+\sigma_2-\sigma_3, -\sigma_1-\sigma_2-\sigma_3 \bigg].
\end{eqnarray}
After projecting onto the low energy subspace of three bands, $\hat{M}_1$ modifies into the following $6 \times 6$ matrix
\begin{eqnarray}
\hat{O}^\prime_3=\frac{1}{\sqrt{3}} \; \bigg[ \sigma_3 \otimes \left(-\frac{1}{3}\; \hat{t}_0-\hat{t}_3 +\frac{1}{\sqrt{3}} \; \hat{t}_8 \right) +\sqrt{2} \sigma_0 \otimes \hat{t}_6 \bigg],
\end{eqnarray}
which leads to
\begin{eqnarray}
U^\dagger \hat{O}^\prime_3 U =\frac{1}{\sqrt{3}} \begin{bmatrix}
-\sigma_3 & \hat{O} & \hat{O} \\
\hat{O} & \sigma_3 & \sqrt{2} \sigma_0 \\
\hat{O} & \sqrt{2} \sigma_0 & -\sigma_3
\end{bmatrix}.
\end{eqnarray}
After projecting out the spin-orbit split band and performing the unitary transformation with $\tilde{U}$ we find that $M_3$ component of SI order parameter couples with the quadratically touching bands as $M_3 \Psi^\dagger \Gamma_{12} \Psi \equiv M_3 \Psi^\dagger \Gamma_3 \Gamma_{45} \Psi$. The other two components of spin ice order couple to underlying fermions according to $M_1 \;\chi^\dagger \hat{O}_{1} \chi$ and $M_2 \;\chi^\dagger \hat{O}_{2} \chi$ where the $8 \times 8$ matrices
\begin{eqnarray}
\hat{O}_1&=&\frac{1}{\sqrt{3}} \; \mathrm{diag} \bigg[ -\sigma_1+\sigma_2-\sigma_3, -\sigma_1+\sigma_2+\sigma_3, -\sigma_1-\sigma_2+\sigma_3, -\sigma_1-\sigma_2-\sigma_3 \bigg], \\
\hat{O}_2&=&\frac{1}{\sqrt{3}} \; \mathrm{diag} \bigg[ -\sigma_1+\sigma_2-\sigma_3, -\sigma_1+\sigma_2+\sigma_3, \sigma_1+\sigma_2-\sigma_3, \sigma_1+\sigma_2+\sigma_3 \bigg].
\end{eqnarray}
Within the $6 \times 6$ subspace we find
\begin{eqnarray}
U^\dagger \hat{O}^\prime_1 U =\frac{1}{\sqrt{3}} \begin{bmatrix}
-\sigma_1 & \hat{O} & -\sqrt{\frac{3}{2}} \sigma_3 \\
\hat{O} & -\sigma_1 & -\frac{i}{\sqrt{2}} \sigma_2 \\
-\sqrt{\frac{3}{2}} \sigma_3 &  \frac{i}{\sqrt{2}} \sigma_2 & -\sigma_1
\end{bmatrix}, \quad
U^\dagger \hat{O}^\prime_2 U =\frac{1}{\sqrt{3}} \begin{bmatrix}
-\sigma_2 & \hat{O} & -i\sqrt{\frac{3}{2}} \sigma_0 \\
\hat{O} & \sigma_2 & -\frac{i}{\sqrt{2}} \sigma_1 \\
i\sqrt{\frac{3}{2}} \sigma_0 &  \frac{i}{\sqrt{2}} \sigma_1 & \sigma_2
\end{bmatrix}.
\end{eqnarray}
Therefore, we conclude that $M_1$ and $M_2$ components of the SI order parameter couple to the quadratically touching bands according to $M_1 \Psi^\dagger \Gamma_{23} \Psi \equiv M_1 \Psi^\dagger \Gamma_1 \Gamma_{45} \Psi$ and $M_2 \Psi^\dagger \Gamma_{31} \Psi \equiv M_1 \Psi^\dagger \Gamma_2 \Gamma_{45} \Psi$ respectively (after roating with $\tilde{U}$). By comparing with Eq.~(\ref{dipole}) and Eq.~(\ref{octupoleT1u}) we conclude that SI order parameter is a particular admixture of dipolar and octupolar quantities following $T_{1u}$ representation.

\emph{3-in 1-out}: A superposition of AIAO and SI order parameters can be captured through $\Psi^\dagger[\varphi \; \Gamma_{45} + \sum_{j=1}^{3} M_j \Gamma_j \Gamma_{45}]\Psi$ and it corresponds to a general form of 3I1O order. For concreteness we can consider a conventional 3I1O configuration that couples with underlying fermions according to $M_1 \;\chi^\dagger \hat{N}_{1} \chi$ where the $8 \times 8$ matrix
\begin{eqnarray}
\hat{N}_1=\frac{1}{\sqrt{3}} \; \mathrm{diag} \bigg[ -\sigma_1+\sigma_2-\sigma_3, -\sigma_1+\sigma_2+\sigma_3, \sigma_1+\sigma_2-\sigma_3, -\sigma_1-\sigma_2-\sigma_3 \bigg]
\end{eqnarray}
After following the steps described above we establish that such 3I1O order couples with the quadratically touching bands according to
\begin{eqnarray}
\frac{1}{2} \; \Psi^\dagger \left[-\Gamma_{45} + \frac{1}{\sqrt{3} } \; \left( \Gamma_{12}+\Gamma_{23}+\Gamma_{31}\right) \right] \Psi,
\end{eqnarray}
which requires $\varphi^2=\mathbf{M}^2$. Therefore, a conventional 3I1O order is an equal superposition of TSI and AIAO orders.

\begin{figure}[htb]
\begin{center}
\includegraphics[scale=0.3]{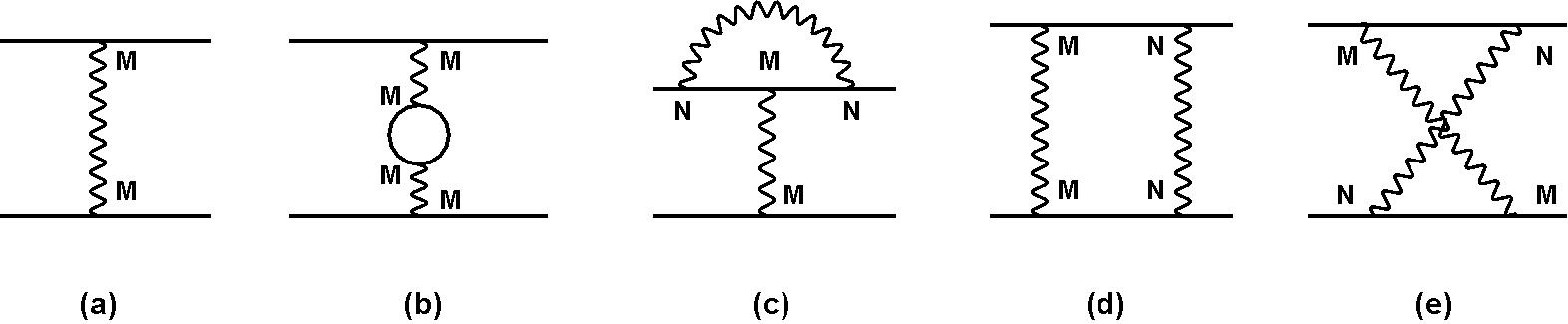}
\caption{Feynmann diagrams needed for the one loop renormalization group calculations for a parabolic semimetal in the presence of generic local interactions as described in Eq.~(\ref{genericint}). In these diagrams $M$ and $N$ stand for $4 \times 4$ hermitian matrices appearing in the interaction vertices as $(\Psi^\dagger M \Psi)^2$ and $(\Psi^\dagger N \Psi)^2$. (a) bare vertex diagram, (b) auto-renormalization of the vertex $(\Psi^\dagger M \Psi)^2$ due to fermion bubble and it is proportional to fermion flavor number, (c) vertex correction to $(\Psi^\dagger M \Psi)^2$ due to $(\Psi^\dagger N \Psi)^2$, (d) ladder diagram and (e) crossed diagram.}
\label{FeynmanDiagram_RG}
\end{center}
\end{figure}

\emph{Ferromagnetism}: The conventional ferromagnetic order couples with the underlying 8-component fermions according to $\mathbf{N} \cdot \; \chi^\dagger \boldsymbol \sigma \otimes \mathbb{1}_{4\times 4} \chi$. Since conventional ferromagnetism corresponds to dipolar ordering, we expect it to couple to four component parabolic fermions according to $\mathbf{N} \cdot \Psi^\dagger \mathbf{J} \Psi$. We demonstrate this for $\mathbf{N}=N_0 \; [1,1,1]/\sqrt{3}$, where the $8 \times 8$ matrix becomes
\begin{eqnarray}
\hat{M}_{FM}=\frac{1}{\sqrt{3}}\: \mathrm{diag} \left[\sigma_1+\sigma_2+\sigma_3, \; \sigma_1+\sigma_2+\sigma_3, \; \sigma_1+\sigma_2+\sigma_3, \; \sigma_1+\sigma_2+\sigma_3 \right].
\end{eqnarray}
By projecting onto the low energy subspace of three bands we obtain
\begin{eqnarray}
\hat{M}^{\prime}_{FM}= S^\dagger \hat{M}_{FM} S=\frac{1}{\sqrt{3}} \left[ \sigma_1+\sigma_2+\sigma_3, \; \sigma_1+\sigma_2+\sigma_3, \;\sigma_1+\sigma_2+\sigma_3 \right].
\end{eqnarray}
After the subsequent unitary transformation with $U$ we find
\begin{eqnarray}
U^\dagger \hat{M}^{\prime}_{FM} U=\frac{1}{\sqrt{3}} \bigg\{
\sigma_0 \otimes \begin{bmatrix}
0 & \frac{1}{\sqrt{3}} & -i \sqrt{\frac{2}{3}} \\
\frac{1}{\sqrt{3}} & 0 & - \frac{2 \sqrt{2}}{3} \\
i \sqrt{\frac{2}{3}} & - \frac{2 \sqrt{2}}{3} & 0
\end{bmatrix}
&+& \sigma_1 \otimes \begin{bmatrix}
0 & 0 & 0 \\
0 & \frac{2}{3} & - i \frac{\sqrt{2}}{3} \\
0 & i \frac{\sqrt{2}}{3} & - \frac{1}{3}
\end{bmatrix}
+ \sigma_2 \otimes \begin{bmatrix}
0 & 0 & 0 \\
0 & \frac{2}{3} & i \frac{\sqrt{2}}{3} \\
0 & -i \frac{\sqrt{2}}{3} & - \frac{1}{3}
\end{bmatrix}
\nonumber \\
&+& \sigma_3 \otimes \begin{bmatrix}
1 & - \frac{i}{\sqrt{3}} & \sqrt{\frac{2}{3}} \\
\frac{i}{\sqrt{3}} & \frac{1}{3} & 0 \\
\sqrt{\frac{2}{3}} & 0 & - \frac{1}{3}
\end{bmatrix}
\bigg\}.
\end{eqnarray}
By performing the final unitary transformation with $\tilde{U}$ we verify our anticipation that the ferromagnetic order displays the dipolar coupling  $2N_0 \; \Psi^{\dagger} \left(J_1+J_2+J_3 \right) \Psi/(3\sqrt{3})$ with quadratically touching bands. \\

\section{Renormalization group flow equations}\label{sec:appendix-RG}

After evaluating the Feynmann diagrams of Fig.~\ref{FeynmanDiagram_RG}, we obtain the following renormalization group flow equations for the six dimensionless coupling constants at one loop order
\allowdisplaybreaks[4]
\begin{eqnarray}
\frac{dg_0}{dl} &=& -\epsilon g_0 + \left[ 3 g_3 g_4+6 g_4 g_5 +6 g^2_5 + g_0g_1+ g_0g_2\right] \\
\frac{dg_1}{dl} &=& -\epsilon g_1 + 8 N g^2_1 + 4 g_1 \left(-g_0 + g_1+2 g_2 -g_3 + g_4 -2 g_5 \right) + \frac{1}{2} \bigg[ g^2_0 + 3 g^2_1 + 2 g^2_2 + g^2_3 + 3 g^2_4 + 6 g^2_5 \nonumber \\
&+& 2 \left(2 g^2_1 +2 g_1 g_2-10 g_1 g_4 -10 g_2 g_5 + 2 g_3 g_5 +4 g_4 g_5 +4 g^2_5 +2 g^2_4 \right) \bigg] \\
\frac{dg_2}{dl} &=& -\epsilon g_2 + 8 N g^2_2 + 4 g_2 \left(-g_0 + 3 g_1 + g_3 -3 g_4 \right) + \frac{1}{2} \bigg[ g^2_0 + 3 g^2_1 + 2 g^2_2 + g^2_3 + 3 g^2_4 + 6 g^2_5 \nonumber \\
&+& 2 \left( 3 g_1 g_2 -15 g_1 g_5 + g^2_2 -5 g_2 g_3 + 3 g_3 g_5 + 6 g_4 g_5 +3 g^2_5 \right) \bigg] \\
\frac{dg_3}{dl} &=& -\epsilon g_3 + 4 N g^2_3 + 2 g_3 \left(-g_0-3 g_1 + 2 g_2 - g_3- 3 g_4 + 6 g_5 \right) +  \bigg[ 3 g_0 g_4 + 3 g_1 g_3 +6 g_1 g_5 -5 g^2_2 \nonumber \\
&+& 2 g_2 g_3 + 6 g_2 g_5-15 g^2_5 \bigg] \\
\frac{dg_4}{dl} &=& -\epsilon g_4 + 4 N g^2_4 + 2 g_4 \left( -g_0 + g_1 -2 g_2-g_3 + g_4 +2 g_5 \right) + \bigg[ g_0 g_3+ 2 g_0 g_5-5 g^2_1 + 7 g_1 g_4 +4 g_1 g_5 \nonumber \\
&+& 2 g_2 g_4 + 4 g_2 g_5-5 g^2_4 -10 g^2_5 \bigg] \\
\frac{dg_5}{dl} &=& -\epsilon g_5 + 4 N g^2_5 + 2 g_5 \left( -g_0 - g_1 + g_3 + g_4 \right) + \bigg[ g_0 g_4+ 2 g_0 g_5-5 g_1 g_2+g_1 g_3+2 g_1 g_4 \nonumber \\
&+& 7 g_1 g_5+ g_2 g_3 + 2 g_2 g_4 + 4 g_2 g_5-5 g_3 g_5-10 g_4 g_5 
\bigg].
\end{eqnarray}
For $N \to \infty$ limit, we recover the mean-field theory predictions for the critical couplings of each ordering channel (decoupled critical points) $g_{1c}=g_{2c}=\epsilon/(8N)$, $g_{3c}=g_{4c}=g_{5c}=\epsilon/(4N)$. How a finite value of $N$ modifies the locations of critical points (no longer decoupled) have to be determined numerically. However, even for moderate values $N \sim 5$ or $6$ the deviation from mean-field locations are not too drastic, due to the control parameter $1/N$.

\begin{figure}[htb]
\begin{center}
\includegraphics[scale=0.3]{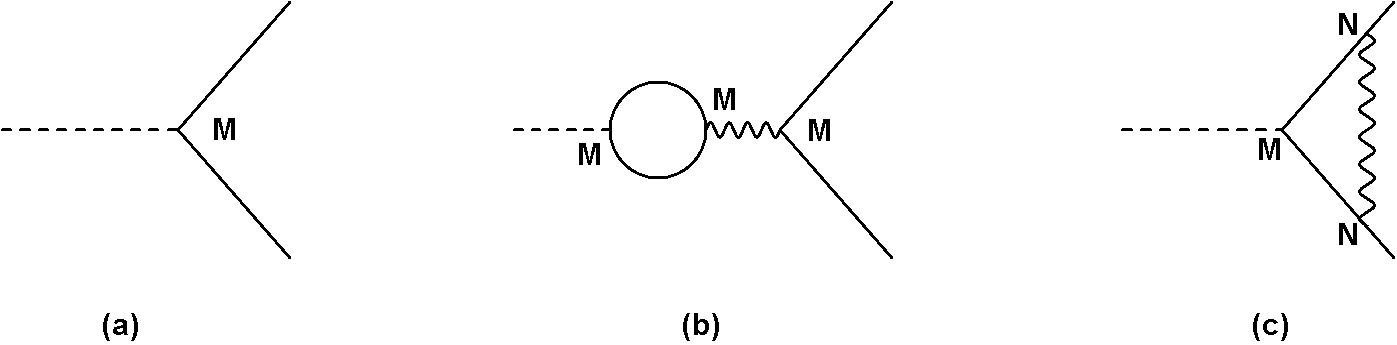}
\caption{The bare susceptibility vertex is shown in panel (a), and its leading order renormalization due to local four-fermion interaction arises from diagrams (b) and (c). Here $M$ and $N$ are $4 \times 4$ hermitian matrices.}
\label{Diagram_RG}
\end{center}
\end{figure}

In order to pin the nature of an emergent broken symmetry phase across a QCP, we need to search for the dominant susceptibility exhibiting strongest divergence. For computing the RG flow of susceptibilities we first couple source fields $\Delta_\mu$ to the fermion bilinears as
\begin{eqnarray}
H_{source}=\int d^3x [\Delta_0 \Psi^\dagger \mathbb{1} \Psi +\Delta_1 \sum_{j=1}^{3} \Psi^\dagger \Gamma_j \Psi + \Delta_2\sum_{j=4}^{5} \Psi^\dagger \Gamma_j \Psi+\Delta_3 \Psi^\dagger \Gamma_{45} \Psi + \Delta_4 \sum_{j=1}^{3} \Psi^\dagger \Gamma_{45} \Gamma_j\Psi+\Delta_5\sum_{j=1}^{3}(\Psi^\dagger \Gamma_{j4} \Psi \nonumber \\ +\Psi^\dagger \Gamma_{j5} \Psi)].
\end{eqnarray}
After evaluating the one-loop diagrams shown in Fig.~\ref{Diagram_RG} we obtain the following RG flow equations for the susceptibilities
\begin{eqnarray}
\frac{d \log \Delta_0}{dl}-2&=&0, \\
\frac{d \log \Delta_1}{dl}-2&=&8 N g_1 +2[-g_0+g_1+2g_2-g_3+g_4-2g_5],\\
\frac{d \log \Delta_2}{dl}-2&=&8 N g_2 +2[-g_0+3g_1+g_3-3g_4],\\
\frac{d \log \Delta_3}{dl}-2&=&4 N g_3 +2[-g_0-3g_1+2g_2-g_3-3g_4+6g_5],\\
\frac{d \log \Delta_4}{dl}-2&=&4 N g_4 +2[-g_0+g_1-2g_2-g_3+g_4+2g_5],\\
\frac{d \log \Delta_5}{dl}-2&=&4 N g_5 +2[-g_0-g_1+g_3+g_4].
\end{eqnarray}  
After solving the RG flow equations for interaction strengths, we also solve for the susceptibilities. When the interaction couplings become relevant (growing with flow time $l$), one or more susceptibilities can diverge. We track the most strongly diverging susceptibility to determine the nature of the emergent ordered state.

\twocolumngrid

\end{document}